\documentclass[fleqn,usenatbib]{mnras}

\usepackage{newtxtext,newtxmath}
\usepackage[T1]{fontenc}

\DeclareRobustCommand{\VAN}[3]{#2}
\let\VANthebibliography\thebibliography
\def\thebibliography{\DeclareRobustCommand{\VAN}[3]{##3}\VANthebibliography}

\usepackage{graphicx}	% Including figure files
\usepackage{amsmath}	% Advanced maths commands

\title[Hot Jupiter formation in dense stellar clusters: A Monte Carlo model applied to 47 Tucanae]{Hot Jupiter formation in dense stellar clusters: A Monte Carlo model applied to 47 Tucanae}

\author[J. A. Wirth et al.]{
James A. Wirth,$^{1}$
Cathie J. Clarke,$^{1}$
Andrew J. Winter,$^{2, 3}$
\\
$^{1}$Institute of Astronomy, University of Cambridge, Madingley Road, Cambridge, CB3 0HA, UK\\ 
$^{2}${Max-Planck Institute for Astronomy (MPIA), Königstuhl 17, 69117 Heidelberg, Germany } \\
$^{3}$Université Côte d'Azur, Observatoire de la Côte d'Azur, CNRS, Laboratoire Lagrange, 06300 Nice, France
}

\date{Accepted XXX. Received YYY; in original form ZZZ}

\pubyear{\the\year{}}

\begin{document}
\label{firstpage}
\pagerange{\pageref{firstpage}--\pageref{lastpage}}
\maketitle

\begin{abstract}
We study the efficiency of high-$e$ migration as a pathway for Hot Jupiter formation in the dense globular cluster 47 Tuc. Gravitational N-body simulations are performed to investigate the orbital evolution of star-planet systems due to dynamical stellar perturbations. Planetary systems that have been scattered into orbits of sufficiently high eccentricity can undergo tidal circularisation, with Hot Jupiter formation being one possible stopping condition. We also account for the possibility of (i) ionisation due to high-energy encounters, (ii) tidal disruption of the planet by tidal forces inside the Roche limit and (iii) Warm Jupiter formation. The orbital evolution of a population of cold Jupiter progenitors, with initial semi-major axes between 1-30 au, is simulated over 12 Gyr using a simplified dynamical model of 47 Tuc. Our computational treatment of dynamical encounters yields an overall HJ occurrence rate of $\mathcal{F}_{\mathrm{HJ}} \approx 5.9 \times 10^{-4}$ per cluster star (a $51$ per cent enhancement relative to the analytic baseline). The probability of Hot Jupiter formation is highest in the core and falls off steeply beyond a few parsecs from the centre of the cluster, where the stellar density is too low to drive efficient eccentricity diffusion. The code can be found here: \url{https://github.com/James-Wirth/HotJupiter}.
\end{abstract}

\begin{keywords}
planets and satellites: formation -- dynamical evolution and stability -- stars: kinematics and dynamics -- globular clusters: individual: 47 Tucanae
\end{keywords}

%%%%%%%%%%%%%%%%%%%%%%%%%%%%%%%%%%%%%%%%%%%%%%%%%%

%%%%%%%%%%%%%%%%% BODY OF PAPER %%%%%%%%%%%%%%%%%%

\section{Introduction}

Hot Jupiters (HJs) are a class of giant exoplanets with short orbital periods $T \lesssim 10$ days and Jovian masses $0.4 \lesssim m/M_{\mathrm{J}} \lesssim 10 $. Interest in these objects was kick-started by the discovery of the very first exoplanet 51 Peg b \citep{MayQue95} -- the prototypical Hot Jupiter -- which hugs its host star in a tight orbit with semi-major axis $a \approx 0.05 \ \mathrm{au}$.

The infeasibility of in-situ HJ formation, due to either gravitational instability \citep{Raf05} or core accretion \citep{Sch14}, suggests that these planets accrete their mass beyond the ice line before migrating inward \citep{Mus15}. There are a number of possible migration pathways, which are broadly categorised into two frameworks: low-$e$ migration (LEM) and high-$e$ migration (HEM). In LEM, gravitational torques due to material in the protoplanetary disk cause the planet to spiral inward on a viscous timescale $\tau_{\nu} \sim \nu^{-1}$. Such a planet may become a HJ if this timescale is substantially shorter than the disk lifetime $\tau_{\mathrm{disk}}$ $\sim$ a few Myr \citep{GolTre80, LinPap86}, and migration can be stopped close to the inner disk edge \citep{Hel19}.

In HEM, high planetary eccentricities are excited by dynamical interactions such as stellar flybys, planet-planet scattering \citep{RasFor96, Car19} and Kozai-Lidov oscillations driven by distant stellar companions \citep{WuMur03}. Interestingly, a large body of observational evidence suggests that HJ hosts preferentially have outer stellar companions \citep{Bel20}. The "Friends of Hot Jupiters" (FOHJ) programme \citep{Knutson14, Ngo16} has shown, via direct-imaging surveys and RV monitoring, that $(47 \pm 7)\%$ of HJ hosts harbour a bound stellar companion between $50-2000 \ \mathrm{au}$ (roughly three times the occurrence rate for field FGK stars). Yet the same study indicates that Kozai forcing by these companions can account for no more than $(16\pm5)\%$ of observed HJs. Additional, non-secular pathways are therefore required, and impulsive encounters with unbound stars offer a natural alternative. There is observational evidence to support this external-perturber channel: Hot Jupiter occurrence is elevated in environments with enhanced stellar density, such as open clusters \citep{Bru16, Win20}, and the encounter cross-section for stellar perturbation varies in direct proportion to the stellar density. In HEM, the planetary eccentricity must be pumped up to extreme values until the planet passes sufficiently close to the host star for tidal torques to shrink the orbit of the planet over the lifetime of the star \citep{Hut1981, Jac08}. However, the tidal forces exerted by the star must not exceed the threshold for planetary disruption \citep{Gui11}. This constraint sets an upper limit on the stellar density, beyond which tidal disruption suppresses HJ formation via HEM \citep{HamTre17, Win22}.

The non-detection of HJs in a Hubble Space Telescope (HST) survey of $\sim$34,000 stars in the dense globular cluster 47 Tuc by \citet{Gil00} was initially surprising and constrained the HJ occurrence rate to $\lesssim 0.2\%$ -- an order of magnitude smaller than the occurrence rate of $(1.2 \pm 0.38) \%$ observed in the field with contemporary RV surveys \citep{Wri12}. It is worth mentioning that stellar multiplicity corrections by \citet{BelKun22} have since reduced the estimated single-star HJ occurrence rate in the field to $(0.96 \pm 0.36)\%$. In a later reassessment of the HST data, \citet{MasWin17} recalculated the expected occurrence rates by assuming that planet statistics in 47 Tuc match those inferred by the Kepler mission \citep{Bor10}. This analysis indicated that the expected number of HJ detections in the HST survey was only $\sim$2 -- far fewer than the $\sim$17 detections initially predicted based on the HJ occurrence rate in the solar neighbourhood. This rendered the null result less statistically significant than previously thought. 

The story is different in open clusters such as M67, for example, where \citet{Bru16} detected 3 HJs within a sample of 66 main-sequence and turnoff stars. This rate ($4.5\%$) is substantially higher than in the field. It is clear that HJ formation in stellar clusters is strongly affected by the local environmental conditions. Notably, HJ formation in the 47 Tuc might be suppressed by its low metallicity, [Fe/H] $\approx -0.7$ \citep{Joh10}. The coagulation of dust grains on the midplane, and their subsequent merging into planetesimals, are both metallicity-dependent processes \citep{Arm10}. The planet-metallicity correlation $\mathcal{F} \sim 10^{1.2[\mathrm{Fe/H}]}$ between giant planet formation rates and metallicity is supported by strong observational evidence \citep{SanIsr01}. It has also been demonstrated that far-ultraviolet (FUV) radiation can drive substantial photo-evaporative mass loss in protoplanetary disks \citep{Haw16}.

High-$e$ migration may also be inefficient in dense cluster environments \citep{HamTre17, Win22}. High rates of stellar perturbation can disrupt the tidal circularisation process, decreasing the likelihood of HJ formation. \citet{Win22} analysed the long-term evolution of planetary eccentricity by deriving a drift-diffusion equation for orbital eccentricity based on the analytic perturbation cross-section \citep{Heggie1996}. By assuming that the initial incidence of gas giants in the range 1-30 au is similar to the field stars in the solar neighbourhood, the authors derived the overall HJ occurrence rate to be $\mathcal{F}_{\mathrm{HJ}} \approx 2.2 \times 10^{-3}$ in 47 Tuc. This estimate is not, however, consistent with recent observational constraints implied by the MISHAPS survey \citep{cri24}. This survey, hereafter C24, used the Dark Energy Camera to search for transit-like eclipses around 19,930 stars in the outskirts of 47 Tuc, but detected no HJs. When combined with the lack of HJ detections in an independent earlier transit search of 34,091 core stars by \citet{Gil00}, the authors obtained a stronger upper bound of $\mathcal{F}_{\mathrm{HJ}, \mathrm{max}} < 1.1 \times 10^{-3}$ for Jovian planets with periods $0.5 \leq T / \mathrm{days} \leq 8.3$ and radii $0.8 < R_{\mathrm{p}} / R_{\mathrm{J}} < 2$.

The significance of the order-unity difference between the analytic predictions and existing observational constraints relies on the accuracy of the former. In particular, there are two key assumptions underpinning the analytic predictions which will be invalid in certain regions of encounter parameter space. The analytic expressions are only correct to quadrupole order (the \textit{tidal} regime) and are derived under the additional assumption that the encounter timescale is large compared to the period of the planetary orbit (the \textit{slow} regime). Inaccuracies in computing perturbation cross-sections in  the regions of encounter parameter space where these assumptions break down could significantly change the overall outcome probabilities, since very small errors in eccentricity can be important for circularisation. A simulation-based study of HJ formation by high-$e$ migration has previously been carried out by \citet{HamTre17} using the regularized restricted three-body code \texttt{RR3}. The authors performed population synthesis calculations of encounters with planetary systems and determined that the HJ occurrence rate peaks at $\sim$2\% at densities $n \sim 4 \times 10^4 \ \mathrm{pc}^{-3}$. This computational approach ensures accuracy in the regions of parameter space where encounters are not tidal and slow, but does not exploit the computational advantage of using the \citet{Heggie1996} expressions in regions of parameter space where the analytic approximation remains valid. For example, a large fraction of the very slowest encounters -- responsible for substantial computational bottlenecks -- occur at large pericentre radii ($R_{\mathrm{peri}}/a_0 \gg 1$), where the analytic approximation for the eccentricity excitation remains valid and can be evaluated in $\mathcal{O}(1)$-time. 

In Section \ref{section:methodology}, we outline our N-body model for eccentricity excitations and investigate the region of parameter space where the analytic approximation breaks down. In Section \ref{section:MonteCarlo_Simulation}, we introduce a `hybrid' Monte Carlo (MC) model which uses direct N-body integration for the subset of perturbing encounters that violate the tidal and slow regime. In Section \ref{section:application}, we apply our model to Monte Carlo simulations of HJ formation in 47 Tuc. Tidal forces due to the host star are implemented using analytic expressions derived in the pseudo-synchronous regime. Planetary systems that reach sufficiently large eccentricities ($e\rightarrow 1$) can quickly circularise to form HJs, provided that they are neither ionised nor tidally disrupted in the process. We run our simulation on an initial population of cold Jupiter progenitors distributed within a simple, time-dependent cluster profile for 47 Tuc and obtain an estimate for the HJ occurrence rate. We give our conclusions in Section \ref{section:conclusions}.

\section{Methodology} \label{section:methodology}

\subsection{Numerical integration of dynamical encounters}

We consider planetary systems with a host star of mass $m_{\star}$ and a planetary companion of mass $m_{\mathrm{p}}$ where $q := m_{\mathrm{p}}/m_{\star} \ll 1$. The two-body gravitational parameter is defined as $\mu := Gm_{\star}(1+q)$. The state of the system prior to a dynamical encounter is a bound Keplerian orbit $\mathbf{r}(t)$ with eccentricity $e$ and semi-major axis $a$. 

The planetary system is shaken up by a series of dynamical perturbations due to passing stars, which follow hyperbolic or parabolic trajectories $\mathbf{R}(t)$ about the host star (since the planet is a test-particle to good approximation). The effect of a dynamical encounter on the planetary orbit is simulated using the \texttt{REBOUND} N-body code \citep{ReiLiu12} and the \texttt{IAS15} integrator \citep{ReiSpe15}. Each encounter is fully parameterised by eleven variables
\begin{equation}
\Pi := (v_{\infty}, b, \Omega, i, \omega, m_{\star}, m_{\mathrm{p}}, m_{\mathrm{pert}}, e, a, \phi)
\end{equation}
where $v_{\infty}$ is the asymptotic relative speed, $b$ is the impact parameter and $(\Omega, i, \omega)$ are the angular orbital elements. The quantity $\phi$ is the initial planetary phase. Unless otherwise specified, we will sample $\Pi$ from the full distributions relevant to 47 Tuc (Table \ref{table:parameter_table}) in our tests and denote this by $\Pi \sim$ \texttt{47Tuc}. Each parameter $\Pi_i$ is sampled via the inverse density $F_i^{-1} : [0,1] \rightarrow \mathbb{R}$, which maps a uniform variable $y \sim U(0,1)$ to a realization $\Pi_i = F_i^{-1}(y)$ according to the underlying distribution. 

The initial planetary separation of cold Jupiter progenitors from their host stars is modelled as an asymmetric broken power-law with a turnover point at $a_{\mathrm{br}} \approx 2.5 \ \mathrm{au}$ \citep{FerMul19}. This reproduces the results of RV surveys at small separations \citep{Cum08}, as well as the drop-off in occurrence rate observed in direct-imaging surveys sensitive from roughly $a \gtrapprox 5 \ \mathrm{au}$.
\begin{equation}
\frac{dN}{d\log a} \propto 
\begin{cases}
\left( a/a_\mathrm{br} \right)^{0.80}, & a \leq a_\mathrm{br} \\
\left( a/a_\mathrm{br} \right)^{-1.83}, & a > a_\mathrm{br}
\end{cases} \quad \text{with } a_\mathrm{br} = 2.5 \ \mathrm{au} \\
\end{equation}
Following \citet{JurTre08}, we adopt a Rayleigh eccentricity distribution with $dF(e) \propto e \exp[-\tfrac{1}{2} (e/\sigma_e)^2]$ with scale parameter $\sigma_e = 0.33$ and truncated at $e_{\mathrm{max}} = 0.6$. We also assume that  the distribution of initial semi-major axes is uniform in $\log a_0$ between 1-30 au, motivated by the findings of \citet{NieDeR19} and the fact that only a few Jupiter mass planets are observed on orbits tighter than $1 \ \mathrm{au}$ \citep{FerMul19}. It is finally worth noting that we have set $m_{\mathrm{p}} = 1 \ M_{\mathrm{J}}$ in all of our experiments.

\begin{table*} 
 \caption{Encounter parameter distributions}
 \label{table:parameter_table}
 \begin{tabular}{llll}
  \hline
  Quantity & Description & Range & Distribution (cumulative density = $F(\cdot)$) \\
  \hline
  \hline
  $v_{\infty}$ & Asymptotic relative speed & $(0, \infty)$ & $dF(v_{\infty}; \sigma_{\mathrm{rel}}) =  \frac{4\pi v_{\infty}^2}{(2\pi \sigma_{\mathrm{rel}}^2)^{3/2}} \exp \left(-\frac{v_{\infty}^2}{2\sigma_{\mathrm{rel}}^2}\right) dv_{\infty}$ \\
  \hline
  $b$ & Impact parameter & $(0, b_{\mathrm{max}}]$ & $F(b) = (b/b_{\mathrm{max}})^2$ \\
  \hline
  $\Omega$ & Longitude of ascending node & $[0, 2\pi)$ & Uniform \\
  $i$ & Inclination & $[0, \pi]$ & $F(i) = \tfrac{1}{2} \sin{i}$\\
  $\omega$ & Argument of periapsis & $[0, 2\pi)$ & Uniform\\
  \hline
  $m_{\star}$ & Host mass & $[0.08 M_{\odot}, 0.8M_{\odot}]$ &  IMF (Equation \ref{equation:imf}) cut off at $m_{\mathrm{br}} = 0.8 M_{\odot}$ \\
  $m_{\mathrm{pert}}$ & Perturbing mass & $[0.08 M_{\odot}, 5 M_{\odot}]$ & IMF (Equation \ref{equation:imf}) \\
  $m_{\mathrm{p}}$ & Planet mass & $M_J$ & \dots \\
  \hline
  $e$ & Initial eccentricity & $[0.05, 0.6]$ & $dF(e) = \tfrac{e}{\sigma_e^2} \mathrm{exp}\left( -\tfrac{e^2}{2\sigma_e^2} \right) de \quad$ with $\sigma_e = 0.33$\\ 
  \hline
  $a$ & Initial semi-major axis & $[1 \ \mathrm{au}, 30 \ \mathrm{au}]$ & 
  $\displaystyle \frac{dN}{d\log a} \propto 
  \begin{cases}
    \left( a/a_\mathrm{br} \right)^{0.80}, & a \leq a_\mathrm{br} \\
    \left( a/a_\mathrm{br} \right)^{-1.83}, & a > a_\mathrm{br}
   \end{cases} \quad \text{with } a_\mathrm{br} = 2.5 \ \mathrm{au}$
    \\
  \hline
  $\phi$ & Initial planetary phase & $[0, 2\pi)$ & $F(\phi) \sim E(\phi) - e\sin{[E(\phi)]}$ \\
  \hline
 \end{tabular}
\end{table*}

\subsubsection{Optimisation}

For non-ionising encounters, the numerical integration can be safely truncated by ignoring the portions of the passing star's trajectory where the ratio $\xi(R) := F(R)/F_{\mathrm{peri}}$ of the perturbing force to its value at pericentre is less than some critical value $\xi_{\mathrm{crit}}$. The quadrupole-order scaling of the perturbing force is $F \sim r/R^3$, but $r(t) \sim a$ remains the same order of magnitude during a non-ionising encounter so the dominant scaling is $\xi(R) \approx (R_{\mathrm{peri}} / R)^3$. 

The boundaries of the numerical integration are taken to be the points where the perturber is at a distance $R_0 = \xi_{\mathrm{crit}}^{-1/3} R_{\mathrm{peri}}$ from the planetary system. This reasoning does not hold for ionising encounters which unbind the planet from the host star. The planetary eccentricity can obtain arbitrarily large values and the integration takes substantially longer to converge. In such cases, however, the final planetary eccentricities are not important for the purposes of our Monte Carlo simulations; ionisation events are considered only as stopping conditions for the dynamical evolution.

\begin{figure} 
\includegraphics[width=\linewidth]{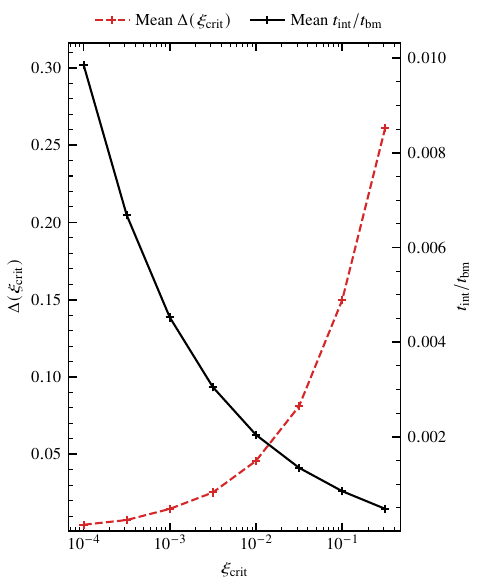}
\caption{The mean error $\Delta(\xi_{\mathrm{crit}})$ due to our truncation of the perturbing orbit at $\xi = \xi_{\mathrm{crit}}$ (Equation \ref{equation:xi}) is shown alongside the mean improvement $t_{\mathrm{int}}/t_{\mathrm{bm}}$ in the integration time. The mean is taken over a randomised sample of $500$ encounters with $\Pi \sim$ \texttt{47Tuc}.}
\label{figure:test_xi}
\end{figure}

We denote the simulated eccentricity excitation for a given encounter obtained with a particular value of $\xi_{\mathrm{crit}}$ by $\epsilon(\xi_{\mathrm{crit}})$, and define the relative error due to our choice by
\begin{equation} \label{equation:xi}
\Delta(\xi_{\mathrm{crit}}) := \bigg{|} \frac{\epsilon(\xi_{\mathrm{crit}}) - \epsilon_{\mathrm{true}}}{\epsilon_{\mathrm{true}}} \bigg{|}, \quad \epsilon_{\mathrm{true}} = \lim_{\xi \rightarrow 0}\epsilon(\xi)
\end{equation}
In practice, we estimate the true value of the eccentricity excitation, $\lim_{\xi\rightarrow 0} \epsilon(\xi)$, by using a `benchmark' value $\xi_{\mathrm{bm}} = 10^{-10}$, corresponding to $R_0 \approx 2000 R_{\mathrm{peri}}$. We adopt a value of $\xi_{\mathrm{crit}} = 10^{-4}$ for the remainder of the work, which yields an average error well below $1\%$ for non-ionising encounters and a substantial 100-fold computation-time improvement compared to the benchmark (Figure \ref{figure:test_xi}). 

\subsubsection{Boundaries of the integration}

\begin{figure*} 
\includegraphics[width=\linewidth]{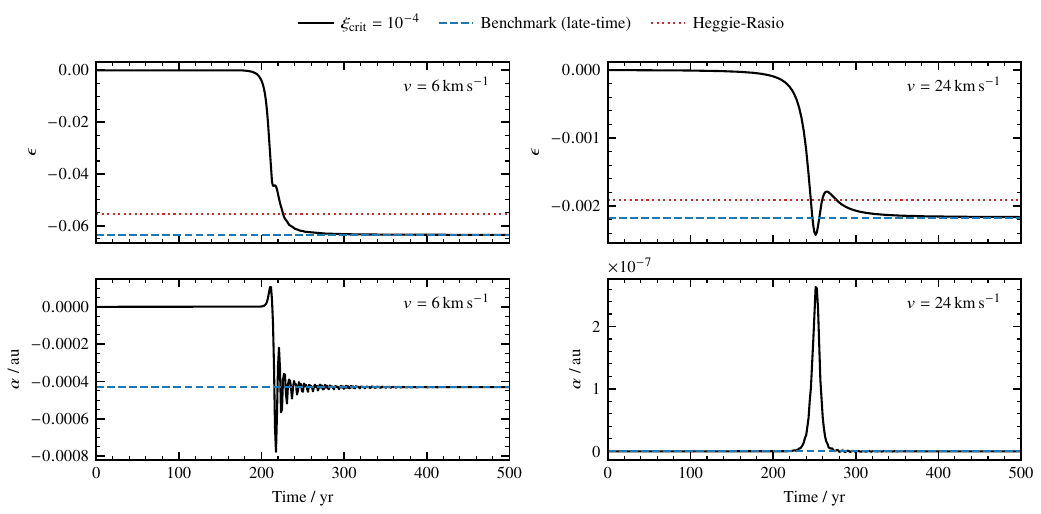}
\caption{The eccentricity excitation $\epsilon$ and semi-major axis excitation $\alpha$ are plotted against time for $v_{\infty} = 6 \ \mathrm{km} \ \mathrm{s}^{-1}$ and $v_{\infty} = 24 \ \mathrm{km} \ \mathrm{s}^{-1}$, with the remaining encounter parameters set to $b = 15 \ \mathrm{au}$, $\Omega = i = \omega = 1 \ \mathrm{rad}$, $e_0 = 0.3$, $a_0 = 1 \ \mathrm{au}$, $m_{\star} = m_{\mathrm{pert}} = M_{\odot}$ and $m_{\mathrm{p}} = M_{\mathrm{J}}$. The simulated results are averaged over the initial planetary phase. The experiments show good convergence to the benchmark value ($\xi_{\mathrm{bm}} = 10^{-10}$, blue) well within the restricted integration time. The encounter with $v = 24 \ \mathrm{km} \ \mathrm{s}^{-1}$ is \textit{secular}, causing a permanent change in eccentricity but no permanent change in semi-major axis.} 
\label{figure:test_time_evolution}
\end{figure*}

The true anomalies $\pm \theta_0$ corresponding to the two points where $R = R_0$ are given by
\begin{equation}
\cos{\theta_0} = e_{\mathrm{pert}}^{-1} \left[ (1+e_{\mathrm{pert}}) \xi_{\mathrm{crit}}^{1/3} - 1\right]
\end{equation}
The perturbing orbit is symmetric under reflection in its apse line, so the amount of time $t_{\mathrm{int}}$ elapsed between $\pm \theta_0$ follows directly from the hyperbolic Kepler equation $M = e\sinh{H} - H$,
\begin{equation} \label{equation:t_int}
t_{\mathrm{int}} = 2(-a_{\mathrm{pert}}^3/\mu)^{1/2} \left[e_{\mathrm{pert}} \sinh{H_0} - H_0\right]
\end{equation}
where the hyperbolic anomaly $H_0$ at the points $\pm \theta_0$ is defined by
\begin{equation}
    \cosh{H_0} = \frac{e_{\mathrm{pert}}+ \cos{\theta_0}}{1 + e_{\mathrm{pert}} \cos{\theta_0}}
\end{equation}
In Figure \ref{figure:test_time_evolution} we illustrate the evolution of the eccentricity and semi-major axis with time over the duration of two example encounters. The encounter with $v_{\infty} = 6 \ \mathrm{km} \ \mathrm{s}^{-1}$ is \textit{non-secular} (resulting in a permanent change in semi-major axis) whereas the encounter with $v_{\infty} = 24 \ \mathrm{km} \ \mathrm{s}^{-1}$ is \textit{secular} (causing no permanent change in semi-major axis). In both cases, the truncated integration with $\xi_{\mathrm{crit}} = 10^{-4}$ exhibits excellent convergence to the benchmark.

\subsection{Parameter space survey} \label{section:parameter_space}

\subsubsection{Analytic approximation}

\begin{figure}
\includegraphics[width=\linewidth]{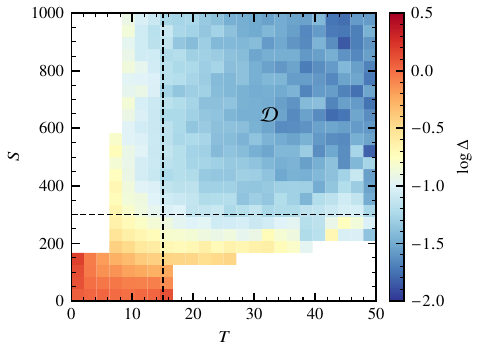}
\caption{The mean binned relative error $\Delta := |(\epsilon_{\mathrm{sim}} - \epsilon_{\mathrm{hr}})/\epsilon_{\mathrm{hr}}|$ of the analytic approximation compared to the simulated values, obtained from a sparse sample of $10^5$ strictly non-ionising encounters with $\Pi \sim$ \texttt{47Tuc}, as a function of $(\mathcal{T}, \mathcal{S})$. We display the results for velocity dispersion $\sigma = 6 \ \mathrm{km} \ \mathrm{s}^{-1}$. The mean relative error is of order $\Delta \lesssim 10\%$ in the region $\mathcal{D}$ where $\mathcal{T} \gtrapprox \mathcal{T}_{\mathrm{min}} = 15$ and $\mathcal{S} \gtrapprox \mathcal{S}_{\mathrm{min}} = 300$.} 
\label{figure:test_params}
\end{figure}

There is an analytic result for the eccentricity excitation of a planetary orbit due to a perturbing encounter \citep{Heggie1996} of the form:
\begin{align}  \label{eq:hr}
\epsilon = \alpha y \left(\frac{a}{R_{\mathrm{peri}}}\right)^{3/2}\{ \Theta_1 \chi + [\Theta_2 + \Theta_3]\psi \}
\end{align}
where the auxiliary functions $\alpha$, $y$, $\chi$, $\psi$ and $\Theta_i$ are defined in Appendix \ref{section:analytic_approximation}. This result is derived under the assumption that the perturbing encounter is both (i) tidal and (ii) slow. The first of these conditions requires that the multipole expansion of the disturbing function $\varphi$ can be safely truncated at quadrupole order ($n=2$), where
\begin{equation} \label{equation:multipole}
\varphi = \frac{Gm_{\mathrm{pert}}}{R} \sum_{n=2}^{\infty} \left( \frac{r}{R} \right)^n P_n (\cos{\theta})
\end{equation}
and the perturbing force is given by $\mathbf{F} = \nabla_{\mathbf{R}} \varphi$. This condition is well satisfied if
\begin{equation} \label{equation:tidal-condition}
\mathcal{T} := R_{\mathrm{peri}}/a \gg 1
\end{equation}
The slowness condition requires that the timescale of the encounter is much greater than that of the planetary orbit. 
\begin{equation} \label{equation:slow-condition}
\mathcal{S} := t_{\mathrm{int}}/t_{\mathrm{per}} \gg 1
\end{equation}
where $t_{\mathrm{per}} = 2\pi a^{3/2}/(Gm_{\star}(1+q))^{1/2}$ is the period of the planetary orbit and $t_{\mathrm{int}}$ is given by Equation \ref{equation:t_int}. It will be useful to define a subset $\mathcal{D}$ of $(\mathcal{T}, \mathcal{S})$-space where the analytic result can be considered a good approximation for the purposes of our Monte Carlo experiment. This will enable us to introduce a "hybrid" model, whereby the full N-body computation is only used in the complement $\mathcal{D}^{\mathrm{C}}$ where the analytic result breaks down. In Figure \ref{figure:test_params} we show the mean binned relative error $\Delta := |(\epsilon_{\mathrm{sim}} - \epsilon_{\mathrm{hr}})/\epsilon_{\mathrm{hr}}|$ of the analytic approximation for a randomised sample of $10^5$ encounters as a function of $(\mathcal{T}, \mathcal{S})$. We have defined
\begin{equation}
\mathcal{D} = \{ (\mathcal{T}, \mathcal{S}) : \mathcal{T} > T_{\mathrm{min}} \ \mathrm{and} \ \mathcal{S} > \mathcal{S}_{\mathrm{min}} \}
\end{equation}
This condition is equivalent to requiring (i) that the $n\geq 3$ terms in the multipole expansion of $\varphi$ are of order $\mathcal{O}(\mathcal{T}_{\mathrm{min}}^{-3})$ and (ii) that the planetary orbit completes a minimum of $\mathcal{S}_{\mathrm{min}}$ full periods during the passage of the perturbing star (between the two points where $\xi = \xi_{\mathrm{crit}}$). We have chosen $\mathcal{T}_{\mathrm{min}} = 15$ and $\mathcal{S}_{\mathrm{min}} = 300$ for the remainder of the work.

\begin{figure} 
\includegraphics[width=\linewidth]{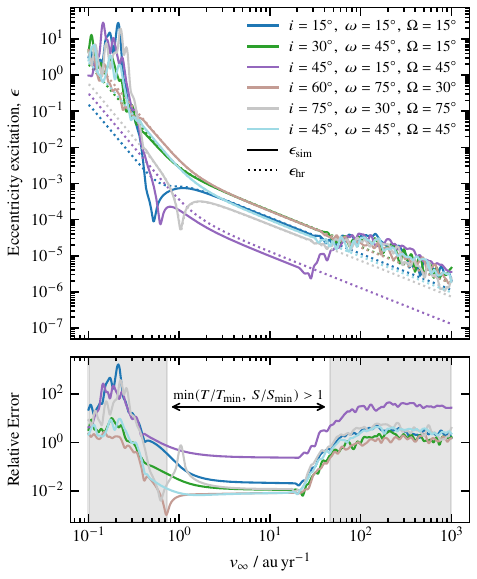}
\caption{The simulated eccentricity excitations ($\epsilon_{\mathrm{sim}}$) are compared against the analytic results ($\epsilon_{\mathrm{sim}}$) over a range of encounter speeds $v_{\infty}/ \mathrm{au} \ \mathrm{yr}^{-1} \in [10^{-1}, 10^{3}]$ and an arbitrary selection of orientations. The remaining free parameters were set to $b = 50 \ \mathrm{au}$, $e_0 = 0.3$, $a_0 = 1 \ \mathrm{au}$, $m_{\star} = m_{\mathrm{pert}} = M_{\odot}$ and $m_{\mathrm{p}} = M_{\mathrm{J}}$. The relative error (lower panel) is small in the region $\mathcal{D}$ of parameter space where $T/T_{\mathrm{min}}>1$ and $S/S_{\mathrm{min}}>1$}.
\label{figure:test_regime}
\end{figure}

In Figure \ref{figure:test_regime} we show the analytic and simulated eccentricity excitation as a function of the relative asymptotic speed $v_{\infty}$ for a range of encounter orientations. The relative error $\Delta := |\epsilon_{\mathrm{sim}} - \epsilon_{\mathrm{hr}}| / |\epsilon_{\mathrm{hr}}|$ for each orientation is displayed in the lower panel. We have indicated the region where both $T/T_{\mathrm{min}} > 1$ and $S/S_{\mathrm{min}} > 1$. At large values of $v_{\infty}$, the slow assumption (\ref{equation:slow-condition}) is violated and the encounter is no longer accurately represented by an interaction between the perturber and an orbit-averaged representation of the planet's position. At small values of $v_{\infty}$ the tidal assumption (\ref{equation:tidal-condition}) also breaks down, indicating that $n > 2$ contributions to the disturbing function become significant.

\subsection{Eccentricity diffusion} \label{section:eccentricity-diffusion}

\subsubsection{Convergence with respect to $b_{\mathrm{max}}$}

\begin{figure} 
\includegraphics[width=\linewidth]{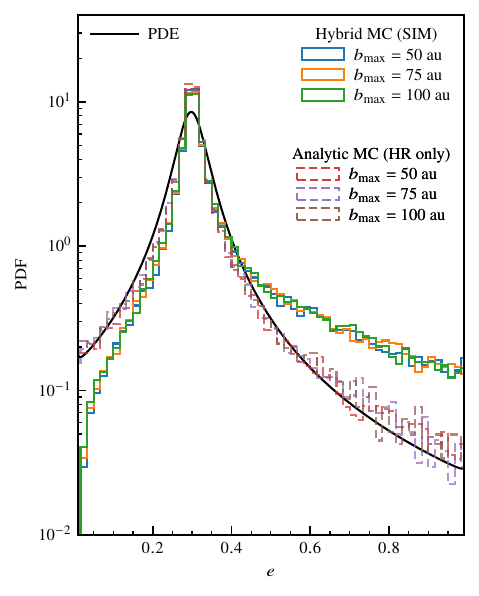}
\caption{The eccentricity distribution of a statistical ensemble of planetary systems after $3$ Gyr of dynamical evolution due to stellar encounters. The initial eccentricity distribution is a sharply peaked Gaussian with $e_0 \sim \mathrm{Normal}(\mu=0.3, \sigma=10^{-3})$ and the remaining free parameters are set to $a_0 = 1 \ \mathrm{au}$, $m_{\star} = m_{\mathrm{pert}} = M_{\odot}$, $m_{\mathrm{p}} = M_{\mathrm{J}}$. The analytic MC curves (dashed histograms) rely solely on the \citet{Heggie1996} approximation, whereas the hybrid MC curves (solid histograms) employ direct N-body simulations in the domain $\mathcal{D}^{\mathrm{C}}$ of parameter space where the approximation breaks down (Section \ref{section:MonteCarlo_Simulation}). The solid black curve is the solution of the PDE in Equation \ref{equation:eccentricity_diffusion}. The MC simulations are computed for $b_{\mathrm{max}}/\mathrm{au} \in \{50,75,100\}$ to demonstrate the good convergence for $b_{\mathrm{max}} \gtrapprox 50 \ \mathrm{au}$. The hybrid MC results exhibit an enhanced rate of eccentricity diffusion toward high eccentricities compared to the analytic prediction.}
\label{figure:try_diffusion}
\end{figure}

The upper limit on the impact parameter has been set to $b_{\mathrm{max}} = 75 \ \mathrm{au}$, which ensures that the encounter sphere is large enough to include almost all encounters of interest, but not too large so as to unnecessarily drive up the computational cost (the mean encounter rate scales as $\beta \sim b_{\mathrm{max}}^2$). In Figure \ref{figure:try_diffusion} we show the distribution of planetary systems in eccentricity space after $3 \ \mathrm{Gyr}$ of diffusion due to randomised encounters, for $b_{\mathrm{max}}/\mathrm{au} \in \{50,75,100\}$. The stepped histograms correspond to Monte Carlo simulations, computed for both (i) purely analytic excitations using the \citet{Heggie1996} expressions and (ii) the hybrid MC scheme, whereby direct N-body integration is employed. In both cases, increasing the radius of the encounter sphere beyond $b_{\mathrm{max}} \gtrapprox 50 \ \mathrm{au}$ has a minimal impact on the diffusion rate, so we have chosen a conservative value of $b_{\mathrm{max}} = 75 \ \mathrm{au}$ for the rest of the work. 

\subsubsection{Comparison with the analytic diffusion rate} \label{section:analytic_model}

The analytic eccentricity distribution (solid black curve in Figure \ref{figure:try_diffusion}) is obtained by solving the drift-diffusion PDE derived by \citet{Win22} under the assumption that all perturbing encounters are hyperbolic. In this regime, the rate of encounters that result in eccentricity excitations $|\epsilon| > \epsilon^{*}$ is given by
\begin{equation}
\Gamma(e, \epsilon^{*}) = \tfrac{1}{\epsilon^{*}} \gamma e \sqrt{1-e^2} 
\end{equation}
where the parameter $\epsilon^{*}$ is the width of each bin in eccentricity space and the constant $\gamma$ controls the rate of diffusion:
\begin{equation} \label{equation:gamma}
\frac{\gamma}{1 \ \mathrm{Myr}} :\approx 0.032 \sqrt{1+q} \frac{n}{10^6 \ \mathrm{pc}^{-3}} \left( \frac{m_{\star}}{1 \ M_{\odot}} \right)^{-1/2} \left( \frac{a_0}{5\ \mathrm{au}} \right)^{3/2}
\end{equation}
We partition the eccentricity space $\mathcal{E} = [0,1)$ for bound orbits into $N$ cells of width $\epsilon^{*}$ with midpoints at $e_i = (i+\frac{1}{2})\epsilon^{*}$ for $i \in \{ 0, 1, \dots, N-1\}$. The rate of change of probability density in the $i^{\mathrm{th}}$ cell is
\begin{equation}  \label{equation:prob-diff-0}
\frac{\partial p(e_i, t)}{\partial t} = -\dot{p}_{\mathrm{out}} + \sum_{j \neq i} \dot{p}_{\mathrm{in}}^{(j)} 
\end{equation}
$\dot{p}_{\mathrm{out}}$ is the probability flux out of this cell and into the neighbouring left-hand and right-hand cells, given by:
\begin{equation}  \label{equation:pflux-out}
\dot{p}_{\mathrm{out}} = \frac{1}{2} p(e_i, t) \bigg[ \Gamma(e_i + \tfrac{1}{2}\epsilon^{*}, \tfrac{1}{2}\epsilon^{*}) + \Gamma(e_i - \tfrac{1}{2}\epsilon^{*}, \tfrac{1}{2}\epsilon^{*}) \bigg]
\end{equation} \label{equation:pflux-in}
$\dot{p}_{\mathrm{in}}^{(j)}$ is the probability flux into this cell from the $j^{\mathrm{th}}$ cell $(j \neq i)$, given by:
\begin{multline}
    \dot{p}_{\mathrm{in}}^{(j)} = \frac{1}{2} p(e_j, t) \bigg[
    \Gamma\Big(e_j - \frac{1}{2} \epsilon^{*} \operatorname{sgn}(j - i), |e_i - e_j| - \frac{1}{2} \epsilon^{*} \Big) \\
     - \Gamma\Big(e_j - \frac{1}{2} \epsilon^{*} \operatorname{sgn}(j - i), |e_i - e_j| + \frac{1}{2} \epsilon^{*} \Big) 
    \bigg]
\end{multline}
In the limit $\epsilon \rightarrow 0$, Equation \ref{equation:prob-diff-0} goes over to a drift-diffusion equation with eccentricity-dependent coefficients,
\begin{equation} \label{equation:eccentricity_diffusion}
\frac{\partial p(e,t)}{\partial t} = \frac{\gamma}{2} \frac{\partial}{\partial e} \left[e\sqrt{1-e^2} \frac{\partial p}{\partial e}\right] - \lim_{\epsilon \rightarrow 0} (\Delta_{+} - \Delta_{-})
\end{equation}
The drift terms $\Delta_{\pm}$ are given by Equation \ref{equation:drift_terms}:
\begin{equation} \label{equation:drift_terms}
\Delta_{\pm} \approx \frac{3\gamma}{4\epsilon^{*}} \bigg[(e\pm \tfrac{3}{2}\epsilon^{*})\sqrt{1-(e\pm \tfrac{3}{2} \epsilon^{*})^2} \frac{\partial p(e\pm \tfrac{3}{2}\epsilon^{*}, t)}{\partial e} \bigg]
\end{equation}

We have solved Equation \ref{equation:eccentricity_diffusion} on $e \in [0,1)$ with a reflective boundary at $e=0$ and an absorbing boundary at $e=1$ (corresponding to ionisation). In Figure \ref{figure:try_diffusion}, the analytic solution of the PDE evolves too slowly toward high eccentricities when compared to the hybrid MC model. The hybrid MC model also exhibits a noticeable upward bias in eccentricity, yielding excess probability density at large $e \approx 1$. This difference is consistent with the approximations underlying the derivation of the PDE in equation \ref{equation:eccentricity_diffusion}, which considers only distant ($T \gg 1$), hyperbolic ($v_{\infty} \gg v_{\mathrm{c}}$) encounters and enforces an equal likelihood of positive and negative eccentricity kicks via the symmetric kernel in Equation \ref{equation:pflux-out}. In dense clusters, gravitational focusing increases the rate of close encounters where the \citet{Heggie1996} approximation breaks down. Direct N-body integrations in this regime often result in larger kicks biased toward higher eccentricity values.

From Figure \ref{figure:try_diffusion} alone, it is not obvious whether the increase in probability density at large eccentricities will drive up HJ formation rates, or instead increase the probability of tidal disruption or ionisation outcomes. For a given semi-major axis $a$, there is only a narrow band of eccentricity values $e \in [e_{\mathrm{min}}, e_{\mathrm{td}}]$ for which the eccentricity is large enough for efficient tidal circularisation to occur ($e \gtrapprox e_{\mathrm{min}}$), but not large enough so as to result in tidal disruption or ionisation ($e < e_{\mathrm{td}} < 1$). An increased diffusion rate would suggest that planetary systems that have reached the critical eccentricity range for HJ formation are also more likely to diffuse out of it before the planetary orbit can circularise. In Section \ref{section:MonteCarlo_Simulation}, we will compute the HJ formation probability explicitly for a typical range of densities, velocity dispersions and initial semi-major axes.

At late times, one might expect the eccentricity distribution to approach the thermalised distribution derived by \citet{Jea19}. The Boltzmann function $f \sim \exp(-E/T)$ of a population of thermalised planetary systems can be written as a function of eccentricity alone, $f(e)de = 2ede$, with all values of $e^2$ equally likely. However, the timescale of thermalisation due to dynamical processing is substantially longer than the typical age of a globular cluster \citep{Gell19}. Eccentricity thermalisation therefore does not play an important role in our simulations.

\section{Monte Carlo simulation of orbital evolution} \label{section:MonteCarlo_Simulation}

The dynamical evolution of the planetary system is simulated with an efficient hybrid approach, whereby eccentricity excitations are calculated using the analytic approximation in the region $\mathcal{D}$ and using our full N-body code otherwise (in $\mathcal{D}^{\mathrm{C}}$). We have implemented parallel processing with \verb|Joblib| by distributing the evaluation of planetary systems across the available CPU cores. The underlying \texttt{IAS15} integrator \citep{ReiSpe15} is a 15th-order, pure-\verb|C99| implementation of Everhart's algorithm. The integration scheme is optimal with respect to Brouwer's law and suitable for long-term orbit integrations. Systematic errors are kept below machine precision. Importantly, \texttt{IAS15} performs well with close encounters; higher-order schemes are often prone to errors due to a lack of sampling points. The wide range of timescales in the problem (i.e. encounter timescales $t_{\mathrm{int}} \sim 100 \ \mathrm{yr}$ compared to the age $t_{\mathrm{age}} \sim 12 \ \mathrm{Gyr}$ of the cluster 47 Tuc) poses a considerable computational challenge. Our hybrid scheme makes it possible to simulate the dynamical evolution of large numbers ($\sim$$10^5$) of planetary systems, over a timescale of several gigayears, in just a few hours on a standard laptop.

\subsection{The dynamical model} \label{section:dynamical_model}

The orbital evolution of the planet is driven by two effects: stochastic kicks from passing stars, and tidal forces from the host star. The latter effect only becomes significant at very large eccentricities $e \approx 1$ (see Equations \ref{equation:tidal-evolution-e} and \ref{equation:tidal-evolution-a}).

\subsubsection{Stochastic kicks}

The planetary system is subject to stochastic perturbations with exponentially distributed waiting times
\begin{equation}
t_{i+1} - t_i \sim \mathrm{Exp}(1/\beta)
\end{equation}
where the rate parameter $\beta(n, \sigma)$ is a function of the local stellar density $n$ and velocity dispersion $\sigma$. We assume that the waiting timescale is much larger than the encounter timescale ($1/\beta \gg t_{\mathrm{int}}$). The mean encounter rate can be derived by considering the flux of passing stars, whose asymptotic relative speed $v_{\infty}$ is sampled from a Maxwell-Boltzmann distribution
\begin{equation} \label{equation:maxwell-boltzmann}
dF(v_{\infty}; \sigma_{\mathrm{rel}}) =  \frac{4\pi v_{\infty}^2}{(2\pi \sigma_{\mathrm{rel}}^2)^{3/2}} \exp \left(-\frac{v_{\infty}^2}{2\sigma_{\mathrm{rel}}^2}\right) dv_{\infty}
\end{equation}
where $\sigma_{\mathrm{rel}} = \sqrt{2} \sigma$ is the local relative velocity dispersion. The flux of encounters with asymptotic relative speed $v_{\infty}$ and impact parameter $b$ is given by $d\beta = (2\pi b db) n v_{\infty} dF$. Integrating over $b \in (0, b_{\mathrm{max}})$ and $v_{\infty} \in (0, \infty)$ yields the mean encounter rate $\beta = 2\sqrt{2\pi} b_{\mathrm{max}}^2 n \sigma_{\mathrm{rel}}$ or more conveniently:
\begin{equation}
\frac{\beta}{1\ \mathrm{Myr}^{-1}} \approx 3.21 \frac{n}{10^{6} \ \mathrm{pc}^{-3}} \left( \frac{b_{\mathrm{max}}}{75 \ \mathrm{au}} \right)^2 \frac{\sigma_{\mathrm{rel}}}{1 \ \mathrm{au} \ \mathrm{yr}^{-1}} 
\end{equation}
The asymptotic relative speed $v_{\infty}$ is sampled from the Maxwell-Boltzmann distribution (Equation \ref{equation:maxwell-boltzmann}). The cumulative density function is analytic and given by
\begin{equation}
F(v_{\infty};\sigma_{\mathrm{rel}}) = \mathrm{erf}\left( \frac{v_{\infty}}{\sqrt{2}\sigma_{\mathrm{rel}}} \right) - \sqrt{\frac{2}{\pi}} \frac{v_{\infty}}{\sigma_{\mathrm{rel}}} \mathrm{exp}\left( - \frac{v_{\infty}^2}{2\sigma_{\mathrm{rel}}^2} \right)
\end{equation}
The impact parameter $b$ is sampled uniformly in $b^2$ up to a maximum radius $b_{\mathrm{max}} = 75 \ \mathrm{au}$. The orientation is sampled isotropically, with $\Omega, \omega \sim U(0, 2\pi)$ and the inclination $i$ sampled uniformly in $\sin{i}$ in $[0, \pi]$. For the perturbing mass $m_{\mathrm{pert}}$ we assume the IMF proposed by \citet{Giersz2011} for 47 Tuc,
\begin{equation} \label{equation:imf}
\xi(m) \sim 
\begin{cases}
m^{-0.4}, & m_{\mathrm{min}} < m < m_{\mathrm{br}} \\
m^{-2.8}, & m_{\mathrm{br}} < m < m_{\mathrm{max}}
\end{cases}
\end{equation}
The function $\xi(m)$ is steep above the turnoff point (with index 2.8 compared to the Salpeter index 2.35), and relatively flat for the lower main sequence (index 0.4). The mass distribution of the host stars is restricted to the main sequence, $m_{\star} < m_{\mathrm{br}} = 0.8 \ M_{\odot}$, and bounded from below by $m_{\mathrm{min}} = 0.08 \ M_{\odot}$ since low mass stars were excluded by the \citet{Gil00} survey. For the perturbing stars, we truncate the distribution at $m_{\mathrm{max}} = 5 \ M_{\odot}$ to exclude short-lived OB stars. This is a minor alteration as only $\sim1\%$ of the original mass distribution due to \citet{Giersz2011} exceeds $5 \ M_{\odot}$.

For encounters computed using the N-body code, the planetary phase at the start of the numerical integration is also randomised. We sample the initial mean anomaly $M$ of the planetary orbit uniformly from the interval $[0, 2\pi)$. The corresponding true anomaly $\theta$ is obtained by solving the Kepler equation $E(\theta) - e \sin{[E(\theta)]} = M$. In the general case this requires numerical methods, but for small eccentricities we can obtain a good approximation by truncating the Fourier expansion (in $\sin{kM}$) at third order in the eccentricity,
\begin{align}
\theta &= M + 2 \sum_{k=1}^{\infty} \left[J_{k}(ke) + \sum_{l=1}^{\infty} \beta^l (J_{k-l}(ke) + J_{k+l}(ke))\right]\frac{\sin{kM}}{k} \\
&\approx M + (2e-\tfrac{1}{4}e^3) \sin{M} + \tfrac{5}{4}e^2 \sin{2M} + \tfrac{13}{12}e^3 \sin{3M} + O(e^4)
\end{align}
where $\beta := e^{-1}(1-\sqrt{1-e^2})$. We make use of this approximation for the marginal speed improvement in cases where $e < 0.3$.

\subsubsection{Tidal forces}

The planet is always subject to tidal forces from the host star. The component of the tidal perturbing force that is perpendicular to the separation vector between the planet and its host star was derived by \citet{Hut1981} and is proportional to $\mathbf{F}_{\perp} \sim -[(\Omega-\dot{\theta})/r^7] \hat{\boldsymbol{\theta}}$, where $\Omega$ and $\dot{\theta}$ are the spin and orbital angular velocities of the planet respectively. In the compact planet regime, the ratio $J_{\mathrm{spin}}/J_{\mathrm{orbit}}$ of the planet's spin and orbital angular momentum must be small. The time-averaged tidal torque $\mathbf{r} \times \mathbf{F}_{\perp} \sim -[(\Omega-\dot{\theta})/r^6]\hat{\mathbf{z}}$ vanishes to good approximation. The circularisation is pseudo-synchronous with $\dot{\Omega} \approx 0$ and the tidal evolution equations for $\dot{e}$ and $\dot{a}$ in \citet{Hut1981} become:
\begin{align} 
\dot{e} &= -\frac{21}{2} k_{\mathrm{p}} \tau_{\mathrm{p}} n^2 q^{-1} \left( \frac{R_{\mathrm{p}}}{a}\right)^5 \frac{e f(e)}{(1-e^2)^{13/2}} \label{equation:tidal-evolution-e} \\
\dot{a} &= -21 k_{\mathrm{p}} \tau_{\mathrm{p}} n^2 q^{-1} \left( \frac{R_{\mathrm{p}}}{a}\right)^5 \frac{ae^2 f(e)}{(1-e^2)^{15/2}} \label{equation:tidal-evolution-a}
\end{align}
where $k_{\mathrm{p}} = 0.25$ is the apsidal motion constant and $n = (\mu/a^3)^{1/2}$ is the mean motion. In this weak-friction model, the tidal time lag $\tau_{\mathrm{p}} = 0.66 \ \mathrm{s}$ accounts for the temporal delay between the maximum amplitude of the tidal bulge and the maximum gravitational deforming potential \citep{Egg98}. We have set the planetary radius $R_{\mathrm{p}} = 0.1 \ \mathrm{R}_{\odot}$. The function $f(e)$ is given by
\begin{equation}
f(e) = \frac{1 + \tfrac{45}{14}e^2 + 8e^4 + \tfrac{685}{224} e^6 + \tfrac{255}{448} e^8 + \tfrac{25}{1792} e^{10}}{1 + 3e^2 + \tfrac{3}{8} e^4}
\end{equation}
It is straightforward to check that Equations \ref{equation:tidal-evolution-e} and \ref{equation:tidal-evolution-a} yield the ODE $da/de = 2ae/(1-e^2)$, which admits the angular momentum $l := a(1-e^2)$ as an integral of motion. The planetary system evolves along the contours of constant $l$ between stochastic kicks.

The expressions for $\dot{e}$ and $\dot{a}$ grow very quickly as $a\rightarrow 0$, which can lead to very poor convergence if not carefully handled. We introduce the dimensionless time $\tau := \beta t$ and choose dynamic integration time-steps $\delta \tau$ such that the maximum fractional change in $e$ or $a$ per time-step is bounded by a regularisation parameter $C$,
\begin{equation}
\mathrm{max}\left(-\frac{d\log{a}}{d\tau}, -\frac{d\log{e}}{d\tau}\right) \delta \tau \leq C
\end{equation}
The parameter value $C := 0.01$ was found to yield good convergence on a relatively fast timescale.

\subsection{Classification of outcomes} \label{outcome_classification}

We classify the final state of each planetary system into one of five distinct target categories:

\begin{enumerate} \label{enumerate:final_states}
    \setlength{\itemsep}{3pt} 
    \setlength{\parskip}{3pt} 
    
    \item \textbf{Ionisation (I)}: The planet becomes unbound from its host star and obtains eccentricity $e \geq 1$. This outcome is only possible when the three-body system has positive total energy, and the ionisation cross-section scales as $\sigma_{\mathrm{ion}} \sim \pi a^2 / v_{\infty}^2$ in the limit of large $v_{\infty}$ \citep{Hut83}. In this regime, the pericentre $R_{\mathrm{peri}}$ of the perturbing orbit must be much smaller than the semi-major axis $a$ of the planetary orbit in order to impart sufficient momentum during the short timescale of the encounter.
    
    \item \textbf{Tidal disruption (TD)}: The planet is scattered into an orbit with closest approach \begin{equation} R_{\mathrm{peri}} = a(1-e) < \eta R_{\mathrm{Roche}} \approx \eta R_{\mathrm{p}} \left( \frac{m_{\star}}{m_{\mathrm{p}}}\right)^{1/3}\end{equation} with $\eta \approx 2.7$ \citep{Gui11}. Inside this radius, tidal forces overcome the self-gravity of the planet, causing the planet to fracture and disintegrate on short timescales. Additionally, asymmetric mass loss may impart sufficient impulse to eject the remaining planetary core from the system. 
    
    \item \textbf{Hot Jupiter (HJ)}: The final period of the planet is less than $10$ days. This occurs when $a < a_{\mathrm{HJ, max}} := (m_{\star}(1+q) T_{\mathrm{HJ, max}}^2)^{1/3}$, where $T_{\mathrm{HJ, max}} := 10 \ \mathrm{days}$. We include in this category those planets with $e > 0$ which have not yet finished circularising. 
    
    \item \textbf{Warm Jupiter (WJ)}: The final period of the planet lies between $10 \ \mathrm{days} < T < 100 \ \mathrm{days}$. 
    
    \item \textbf{No Migration (NM)}: The remaining portion of phase space where the final period of the planet is larger than $100$ days, and the planet has neither been ionised nor has it breached $r = R_{\mathrm{td}}$.
\end{enumerate}

\subsubsection{Stopping conditions} \label{section:stopping_conditions}

Planets that have been ionised or tidally disrupted at an intermediate time $T_{\mathrm{stop}}$ are no longer evolved beyond this point. We have also defined the stopping time for HJs to be the first time at which both $a < a_{\mathrm{HJ, max}}$ and $e < 10^{-3}$. It is acceptable to cease evolution beyond this point, because the scaling $\epsilon \sim a^{3/2}$ of eccentricity perturbations (Equation \ref{equation:eccentricity_excitation}) implies that there is only a very small probability that a HJ is scattered out of its orbit in the remaining time.

\begin{figure*} 
\includegraphics[width=\linewidth]{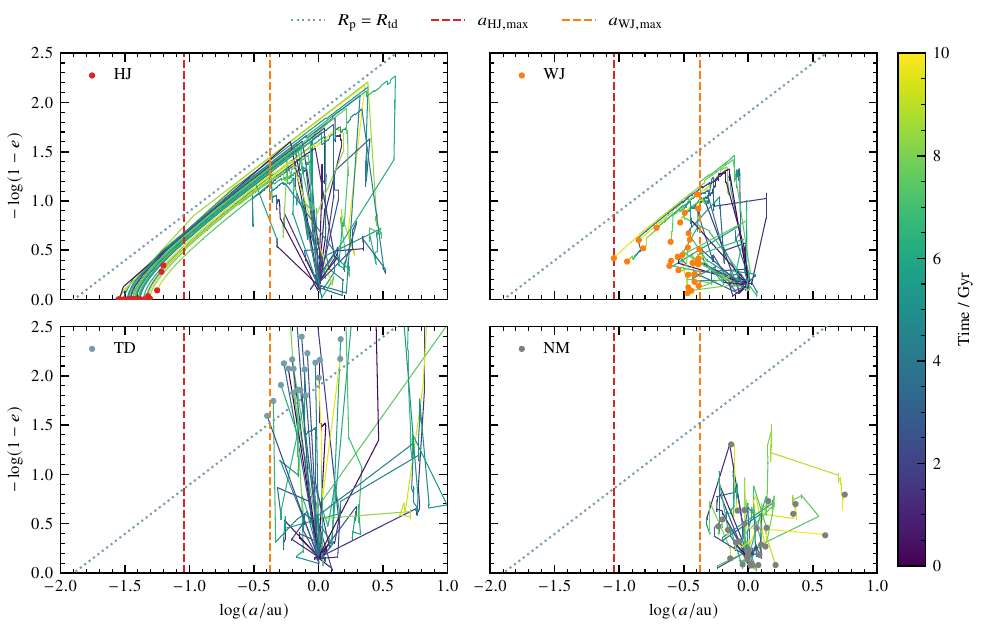}
\caption{The paths in $(a, \tfrac{1}{1-e})$-space for 30 simulated outcomes in each of the categories: Hot Jupiter (HJ, top left), Warm Jupiter (WJ, top right), Tidal Disruption (TD, lower left), No Migration (NM, lower right). In each example, the initial eccentricity is $e_0 = 0.3$ and the initial semi-major axis is $1 \ \mathrm{au}$. The systems are evolved for a maximum of $t = 10 \ \mathrm{Gyr}$ in a local environment with $n = \mathrm{10^4} \ \mathrm{pc}^{-3}$ and $\sigma = 6 \ \mathrm{km} \ \mathrm{s}^{-1}$. The dotted blue line $(R_{\mathrm{peri}} = R_{\mathrm{td}})$ is the boundary of the top-left region where the planetary system is tidally destroyed.} 
\label{figure:show_paths}
\end{figure*}

In Figure \ref{figure:show_paths} we show a representative sample of paths in $(a, \tfrac{1}{1-e})$-space for HJ, WJ, TD and NM outcomes. The majority of HJs are fully circularised ($e \lesssim 10^{-3}$) after $10 \ \mathrm{Gyr}$ however we still consider nearly-circularised planets with $T < 10 \ \mathrm{days}$ to be HJs. The formation pathways of all observed HJs involve at least one significant perturbation that scatters the planet into a highly eccentric orbit within a single timestep. Most such planets then have sufficient time to undergo complete tidal circularization before the end of the simulation. A large fraction of observed WJs, on the other hand, were formed via a direct perturbation into the WJ-region in phase space (bypassing the typical pathway of eccentricity diffusion followed by tidal circularization along the contours of $l$, which is exhibited by HJs). 

\subsubsection{Analytic pipeline} \label{section:analytic_pipeline}

The outcome probabilities can be approximated analytically using the theoretical pipeline introduced by \citet{Win22}. 
\begin{equation}
P_{\{\mathrm{tide, ion}\}} = \frac{\Gamma_{\{\mathrm{tide, ion}\}}}{\Gamma_{\mathrm{tide}} + \Gamma_{\mathrm{ion}}} [ 1 - e^{-(\Gamma_{\mathrm{ion}} + \Gamma_{\mathrm{tide}})t}]
\end{equation}
In these equations, $\Gamma_{\mathrm{ion}}$ and $\Gamma_{\mathrm{tide}}$ are the ionisation rate and tidal rate respectively. The tidal rate is the lumped probability that a planet is influenced by tidal dissipation in some capacity (thereby including both HJ formation and tidal disruption outcomes). It can be expressed analytically by Equation \ref{equation:tidal_rate},
\begin{equation} \label{equation:tidal_rate}
\Gamma_{\mathrm{tide}} = \frac{\gamma e_0 \sqrt{1-e_0}}{2\left[\mathrm{min}(\mathrm{max}(e_{\mathrm{age}}, e_{\mathrm{min}}), e_{\mathrm{td}})-e_0 \right]}
\end{equation}
where $e_{\mathrm{age}}$ is the eccentricity at which the circularisation timescale equals the cluster age, $e_{\mathrm{min}}$ is the eccentricity at which tidal forces balance the perturbative effect of stellar encounters, and $e_{\mathrm{td}}$ is the eccentricity above which planets are tidally disrupted. The rate parameter $\gamma$ is given by Equation \ref{equation:gamma}.
The ionisation rate is obtained by feeding the scattering cross-section $\sigma_{\mathrm{ion}} \sim \pi a^2 / v_{\infty}^2$, derived by \citet{HutBah83} in the close encounter limit, into the differential expression for the ionisation rate,
\begin{equation}
d\Gamma_{\mathrm{ion}} = n\sigma_{\mathrm{ion}} v_{\infty} F(v_{\infty};\sigma_{\mathrm{rel}}) \xi(m_{\mathrm{pert}}) dv_{\infty} dm_{\mathrm{pert}}
\end{equation}
This exercise yields the following expression:
\begin{equation}
\frac{\Gamma_{\mathrm{ion}}}{1 \ \mathrm{Myr}^{-1}} \approx 0.028 \mathcal{M}_{\star}^{(\mathrm{ion})} \left( \frac{\sigma}{10 \ \mathrm{km} \ \mathrm{s}^{-1}} \right)^{-1} \frac{m_{\star}}{1 \ M_{\odot}} \frac{a}{5 \ \mathrm{au}} \frac{n}{10^6 \ \mathrm{pc}^{-3}}
\end{equation}
where the term $\mathcal{M}_{\star}^{(\mathrm{ion})} := \int_0^{\infty} dm_{\mathrm{pert}}(1+q_{\mathrm{pert}} + q) q_{\mathrm{pert}}^{1/3} \xi(m_{\mathrm{pert}})
$ captures the dependence on the initial mass function and $q_{\mathrm{pert}} := m_{\mathrm{pert}}/(m_{\star}(1+q))$. Furthermore, the probabilities for HJ formation and tidal disruption are given by:
\begin{align}
P_{\mathrm{HJ}} &= (1-\lambda_{\mathrm{td}}) P_{\mathrm{tide}} \\
P_{\mathrm{TD}} &= \lambda_{\mathrm{td}} P_{\mathrm{tide}}
\end{align}
The quantity $\lambda_{\mathrm{td}}$ is the fraction of circularising planets that are tidally destroyed, 
\begin{equation}
\lambda_{\mathrm{td}} = 1 - \exp\left( -\frac{R_{\mathrm{td}}}{r_{\mathrm{p, max}}}\right)
\end{equation}
with $r_{\mathrm{p, max}} = l_{\mathrm{max}}/(1+e_0)$  being the maximum circularisation radius for which tidal forces balance the perturbative effect of stellar encounters (Equation 47 of \citet{Win22}).

\subsection{Results of the simulation and comparison with analytic predictions}
\label{section:results_of_simulations}

\begin{figure*} 
\includegraphics[width=\linewidth]{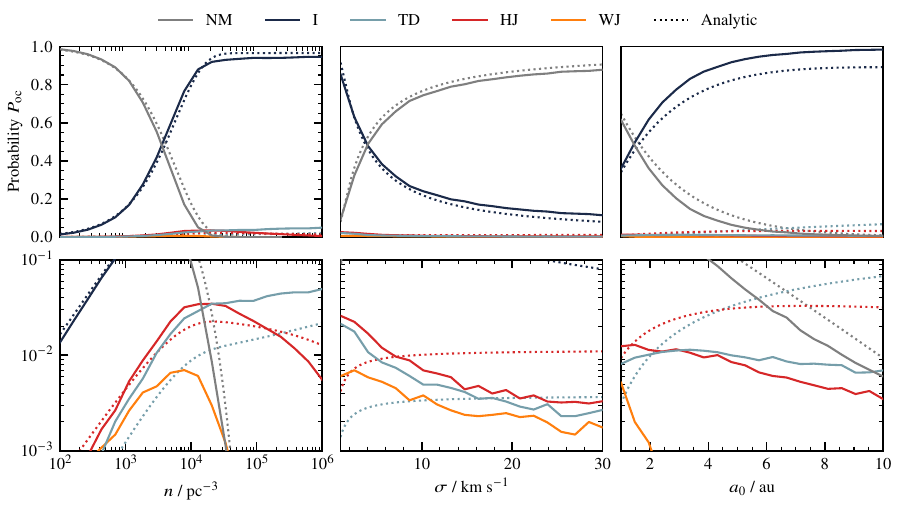}
\caption{The simulated probability for each outcome listed in Section \ref{outcome_classification} is plotted against (i) the local stellar density $n \in [10^{2} \ \mathrm{pc}^{-3}, 10^{6} \ \mathrm{pc}^{-3}]$, (ii) the initial semi-major axis $a_0 \in [1 \ \mathrm{au}, 30 \ \mathrm{au}]$, and (iii) the local velocity dispersion $\sigma \in [1 \ \mathrm{km} \ \mathrm{s}^{-1}, 30 \ \mathrm{km} \ \mathrm{s}^{-1}]$. The canonical parameter values used in each plot, unless the parameter in question is being varied, are $n = 5\times 10^{3} \ \mathrm{pc}^{-3}$, $\sigma = 6 \ \mathrm{km} \ \mathrm{s}^{-1}$, $a_0 = 1 \ \mathrm{au}$, $m_{\star} = M_{\odot}$, and $m_{\mathrm{p}} = M_{\mathrm{J}}$. The initial eccentricity $e_0$ is drawn from a Rayleigh distribution with $\sigma_e = 0.33$ and cut off at $e = 0.6$. The experiments are evolved for $10 \ \mathrm{Gyr}$. The analytic results are also shown (dotted lines).} 
\label{figure:test_model_param_plot}
\end{figure*}

The dependence of the outcome probabilities $P_{\mathrm{oc}}$ on the local stellar density $n$, the velocity dispersion $\sigma$ and the initial semi-major axis are shown in Figure \ref{figure:test_model_param_plot}. The simulated (hybrid-model) results are displayed alongside the analytic predictions described in Section \ref{section:analytic_pipeline}.

The simulated HJ formation rate exceeds the analytic prediction between $10^{3} \ \mathrm{pc}^{-3}$ and $10^{5} \ \mathrm{pc}^{-3}$, peaking at $P_{\mathrm{HJ}} \sim 2 \%$ at an intermediate density of $n \sim 10^4 \ \mathrm{pc}^{-3}$ for the fiducial values $\sigma = 6 \ \mathrm{km} \ \mathrm{s^{-1}}$ and $a_0 = 1 \ \mathrm{au}$. This is in close agreement with the previous numerical experiments of \citet{HamTre17}. We attribute this increase in $P_{\mathrm{HJ}}$ to the enhanced rate of eccentricity diffusion observed in Figure \ref{figure:try_diffusion}. The rate of tidal disruption overtakes the rate of HJ formation beyond $n \sim 10^4 \ \mathrm{pc}^{-3}$, since very high rates of stellar perturbation increase the likelihood of HJ candidates being scattered out of their orbits during tidal circularisation. WJ formation is the least likely of all outcomes. A large fraction of WJs are produced by non-secular perturbations which bring the semi-major axis to below the critical radius $a_{\mathrm{WJ, max}}$ in one timestep (without having to go through the eccentricity diffusion + tidal circularisation process). This explains the steep decline in WJ formation beyond $a_0 \gtrapprox 2 \ \mathrm{au}$, where the probability of non-secular encounters that can scatter the planet into a WJ orbit decreases.

The simulated probability of both HJ formation and tidal disruption is suppressed by a factor of order unity relative to the analytic estimates at velocity dispersions $\sigma \gtrapprox 7.5 \ \mathrm{km} \ \mathrm{s}^{-1}$ and at initial semi-major axes $a_0 \gtrapprox 1 \ \mathrm{au}$. This is balanced by an increase in ionisation events, indicating that encounters with large relative velocities and large host-planet separations are more likely to unbind the planet ($e > 1$) than to excite it into a critical eccentricity bin ($e \in [e_{\mathrm{min}}, e_{\mathrm{td}}]$) where it may begin to circularise. Encounters involving planets on wide orbits with large semi-major axes are likely to violate the tidal assumption, since $\mathcal{T} \sim a^{-1}$. Terms of order $n>2$ in the multipole expansion (Equation \ref{equation:multipole}) are expected to become important and contribute to the discrepancy. Similarly, encounters with large asymptotic relative speeds are likely to violate the slow assumption. This can be seen as follows: the angular speed of the perturbing star at pericentre satisfies $\dot{\theta}^2 = Gm_{\mathrm{tot}}(1+e_{\mathrm{pert}})/R_{\mathrm{peri}}^3$ and the perturbing eccentricity satisfies $e_{\mathrm{pert}}^2 = 1 + b^2 v_{\infty}^4 / \mu^2$. The timescale of the encounter, $\mathcal{S} \propto 1/\dot{\theta}$, approaches zero for very large $v_{\infty}$.

\section{Application to 47 Tuc} \label{section:application}

The structural and kinematical properties of the globular cluster 47 Tuc have been extensively studied through photometric, spectroscopic and proper motion surveys \citep{Mey88, Tra95, Lan10}. Dynamical models range from single-mass and multi-mass King profiles \citep{DaCosta85} to more sophisticated Fokker-Planck and Monte Carlo simulations \citep{Giersz2011}, which account for mass segregation, stellar evolution and binary interactions. Despite the wealth of observational data, significant uncertainties remain in the cluster's dynamical history. 

\subsection{Toy model} \label{section:toy_model}

In this work, we do not attempt to capture the full dynamical complexity of 47 Tuc in our cluster model. Instead, we introduce a simplified toy model -- neglecting effects such as dynamical mixing -- to provide a suitable background on which to test the hybrid MC model against the analytic MC model. 

The results of the full-scale Monte Carlo modelling of 47 Tuc by \citet{Giersz2011} are particularly useful in this context. The initial and present-day values of key cluster parameters in their `Model A' are summarised in Table \ref{table:47_tuc_parameters}. Given that the tidal radius $r_{\mathrm{t}}$ is always very large compared to the half-mass radius $r_{\mathrm{h}}$, the cluster can be approximated as evolving purely by dynamical relaxation. The half-mass relaxation timescale $t_{\mathrm{h}}$ is given by:
\begin{equation}
t_{\mathrm{h}} \sim \frac{N}{\ln{\gamma N}} \left( \frac{r_{\mathrm{h}}^3}{GM}  \right)^{1/2}
\end{equation}
where $\gamma$ is a constant and $\ln{\gamma N}$ is the Coulomb logarithm. On the evolutionary track where the cluster evolves by pure relaxation, the half-mass relaxation timescale $t_{\mathrm{h}}$ scales with the age of the cluster, $t$. We therefore model the evolution of $r_{\mathrm{h}}(t)$ by a function
\begin{equation}
r_{\mathrm{h}}(t) = (r_{\mathrm{h}}(0)^{3/2} + At)^{2/3}
\end{equation}
where the constant $A \approx 6.99 \times 10^{-4} \ \mathrm{pc}^{3/2} \mathrm{Myr}^{-1}$ is fitted to the values of $M$ at $t= 0 \ \mathrm{Gyr}$ and $t= 12 \ \mathrm{Gyr}$ from Table \ref{table:47_tuc_parameters}. This expression reproduces the $r_{\mathrm{hm}}(t) \sim t^{2/3}$ scaling in the relaxation dominated regime. Since the stellar population size only decreases by $7.5\%$ over $12 \ \mathrm{Gyr}$, we assume a constant total population of $N = 2 \times 10^6$ stars in our simulation. We also assume a constant mass function (see Equation \ref{equation:imf}), with mean stellar mass $\langle m \rangle \approx 0.8 \ M_{\odot}$, and neglect mass loss due to stellar evolution. The number density of this population is modelled using a Plummer-sphere,

\begin{equation}
n(r, t) = \frac{3N}{4\pi a(t)^3} \left( 1 + \frac{r^2}{a(t)^2} \right)^{-5/2} \label{equation:rho}
\end{equation}

where the scaling constant $a(t)$ can be expressed in terms of the half-mass radius by $a(t) \approx 0.766\ r_{\mathrm{h}}(t)$. 

Although we have neglected mass loss in our treatment of the simulated stellar population, we model the background gravitational potential as being sourced by a time-dependent total mass $M_{\mathrm{dyn}}(t)$. For simplicity, we assume a linear rate of mass loss between $M(t/\mathrm{Gyr} = 0)$ and $M(t/\mathrm{Gyr} = 12)$. This time-dependent mass $M_{\mathrm{dyn}}(t)$ reflects the cumulative effects of stellar evolution, tidal stripping and dynamical relaxation in dynamical quantities like the velocity dispersion, even though no stars are actually removed from the simulated sample. The Plummer velocity-dispersion is then given by:

\begin{align}
\sigma^2(r, t) &= \frac{GM_{\mathrm{dyn}}(t)}{6\sqrt{r^2 + a(t)^2}} \label{equation:sigma}
\end{align}

\begin{figure} 
\includegraphics[width=\linewidth]{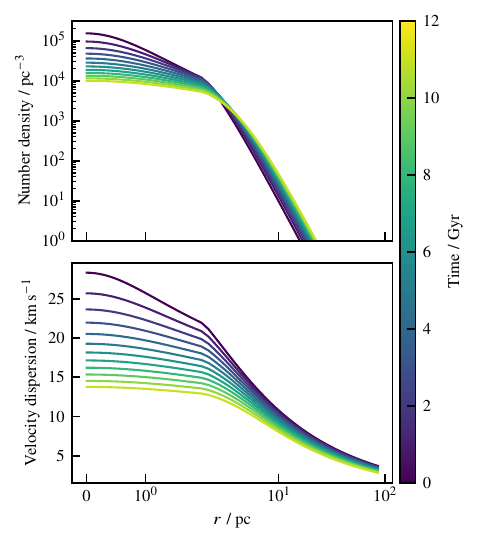}
\caption{The time-evolution of the local stellar density $n(r, t)$ (Equation \ref{equation:rho}) and the velocity dispersion $\sigma(r, t)$ (Equation \ref{equation:sigma}) for the toy model of 47 Tuc presented in Section \ref{section:toy_model}.}
\label{figure:test_density_evolution}
\end{figure}

The time-evolution of $n(r,t)$ and $\sigma(r,t)$ is shown in Figure \ref{figure:test_density_evolution}. For the purposes of our Monte Carlo simulation, we consider a statistical ensemble of $N_{\mathrm{MC}} = 500,000$ planetary systems\footnote{The value of $N_{\mathrm{MC}}$ is set purely by computational constraints. The results of the Monte Carlo experiment approach the true values as $N_{\mathrm{MC}} \rightarrow \infty$.}, each containing a host star and a cold Jupiter progenitor at an initial orbital separation $a_0$ between 1-30au. The radial coordinates $r$ of the planetary systems, measured from the centre of the cluster, are distributed in proportion to the density profile: $f(r, t) \sim 4\pi r^2 \rho(r, t)$. The \textit{Lagrangian} radius of each planetary system is held fixed during the simulation as the cluster expands. This approach neglects relaxation effects that can change the mean Lagrangian radius of individual systems, and furthermore does not model the stellar distribution function (and therefore cannot follow variations in density around the cluster orbit). 

\begin{table} \
 \caption{Initial and present-day conditions for 47 Tuc as given by `Model A' of \citet{Giersz2011}.}
 \label{table:47_tuc_parameters}
 \begin{tabular}{lll}
  \hline
  Parameter & $t = 0 \ \mathrm{Gyr}$ & $t = 12 \ \mathrm{Gyr}$ \\
  \hline
  \hline
  $N$ & $2.00 \times 10^6$ & $1.85 \times 10^6$ \\
  \hline
  $M \ / \ M_{\odot}$ & $1.64 \times 10^6$ & $0.90 \times 10^6$ \\
  \hline
  $r_{\mathrm{t}} \ / \ \mathrm{pc}$ & $86$ & $70$\\
  \hline
  $r_{\mathrm{h}} \ / \ \mathrm{pc}$ & $1.91$ & $4.96$  \\
  \hline
  \hline
 \end{tabular}
\end{table}

\subsection{Results}

\subsubsection{Overall per-star outcomes}

We performed Monte Carlo simulations of the eccentricity evolution of planetary systems with (i) an analytic MC model and (ii) a hybrid MC model (which includes simulated encounters, as explained in Section \ref{section:MonteCarlo_Simulation}). To isolate the effect of our new treatment of encounters, we ran both simulations on the same cluster model (Section \ref{section:toy_model}). The probabilities for each final state (listed in Section \ref{enumerate:final_states}) are summarised in Table \ref{table:47_tuc_results}. These probabilities are conditioned on the assumption that the star hosts a viable HJ candidate, with a mass of order $M_{\mathrm{J}}$ and an initial semi-major axis $a_0$ drawn from a broken power-law distribution between 1-30 au. 

With the hybrid MC model, Hot Jupiters form at a rate of $P_{\mathrm{HJ}} \approx 5.9 \times 10^{-3}$: an increase of roughly 51 per cent compared to the analytic MC model ($P_{\mathrm{HJ}} \approx 3.9 \times 10^{-3}$). This effect can be attributed to the enhanced rate of eccentricity diffusion of simulated planetary systems compared to analytic predictions (see Figure \ref{figure:try_diffusion}). Dynamical encounters in the core of 47 Tuc are much more strongly interacting than implied by the \citet{Heggie1996} expressions, which ignore octopole contributions to the disturbing force and assume that all encounters are slow. We also observe a large increase in the ionisation probability, from $P_{\mathrm{ion}} \approx 0.22$ to $P_{\mathrm{ion}} \approx 0.62$, which is accompanied by a reduction in no-migration outcomes. WJ formation is by far the rarest outcome in both models. WJs occur either as an `intermediate' state, where insufficient time has elapsed for the semi-major axis to shrink to $a < a_{\mathrm{HJ, max}}$ during tidal circularisation, or alternatively by non-secular scattering directly into the relevant region of $(a,e)$ phase space. Tidal disruption occurs in just under $1\%$ of our hybrid-model experiments.

\begin{table*}
 \caption{The overall outcome probabilities for stars hosting HJ candidates, obtained from the hybrid MC model for 47 Tuc with $N_{\mathrm{MC}} = 500,000$. The Poisson error in the overall probabilities is $\Delta_{\mathrm{MC}} \approx 0.0014$. We also show the conditional outcome probabilities for planetary systems with present-day radii with $r<0.5 \ \mathrm{pc}$ (core stars only) and $r> 8 \ \mathrm{pc}$ (outskirts stars only).}
 \label{table:47_tuc_results}
 \begin{tabular}{l|ccc|ccc}
  \hline
  Outcome & \multicolumn{3}{c|}{Analytic $P_{\mathrm{oc}}$} & \multicolumn{3}{c}{Hybrid $P_{\mathrm{oc}}$} \\
          & $r<0.5$ pc & $r>8$ pc & Overall & $r<0.5$ pc & $r>8$ pc & Overall \\
  \hline\hline
  No-migration      & 0.5380 & 0.9644 & 0.7706 & 0.0113 & 0.8779 & 0.3635 \\
  Ionisation        & 0.4460 & 0.0350 & 0.2196 & 0.9480 & 0.1210 & 0.6199 \\
  Tidal Disruption  & 0.0120 & 0.0002 & 0.0060 & 0.0280 & 0.0005 & 0.0106 \\
  Hot Jupiter       & 0.0040 & 0.0004 & 0.0039 & 0.0120 & 0.0004 & 0.0059 \\
  Warm Jupiter      & 0      & 0      & 4E-5   & 0.0007 & 7E-5   & 2E-4 \\
  \hline\hline
\end{tabular}
\end{table*}

\subsubsection{Statistical distribution of stopping time, cluster radius and semi-major axis}

\begin{figure*} 
\includegraphics[width=\linewidth]{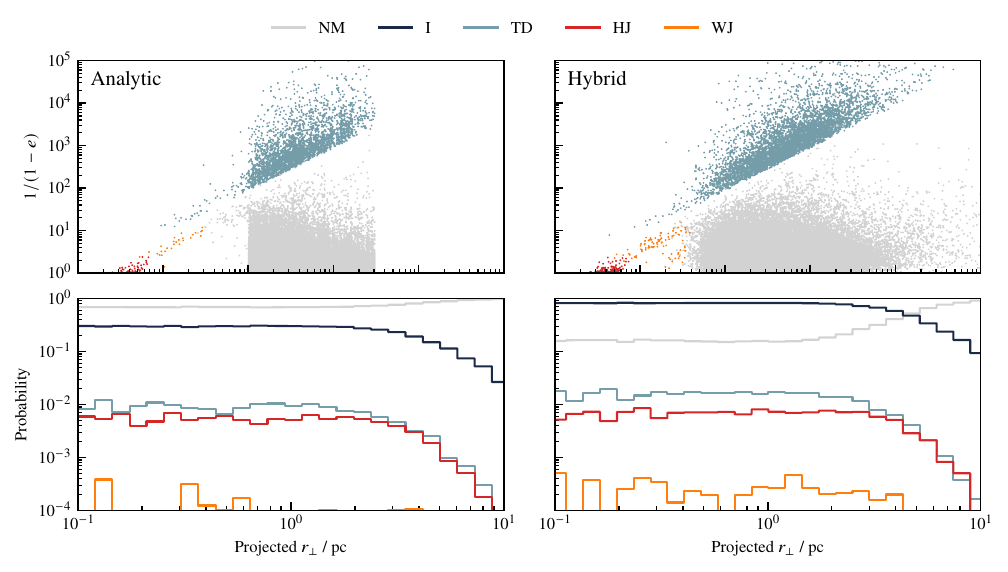}
\caption{A comparison of the outcomes from the analytic and hybrid MC models applied to the dynamical model of 47 Tuc. \textbf{Upper}: The distribution of outcomes in the $(a, \tfrac{1}{1-e})$-plane. (There is some bleed between the HJ, WJ and NM regions since the critical semi-major axes $a_{\mathrm{HJ, max}}$ and $a_{\mathrm{WJ, max}}$, corresponding to $T_{\mathrm{HJ, max}} = 10 \ \mathrm{days}$ and $T_{\mathrm{WJ, max}} = 100 \ \mathrm{days}$ respectively, are functions of the host mass $m_{\star}$ which is randomised for each system.) \textbf{Lower}: The outcome probability $P_{\mathrm{oc}}$ as a function of the projected radius $r_{\perp}$ from the centre of the cluster. The plot has been clipped to show only the bins containing over $10^3$ experiments, so that the upper bound in the Poisson error is $\sim 0.03$.}
\label{figure:comparison_fig}
\end{figure*}

The distribution of final states in the $(a, \tfrac{1}{1-e})$ plane is illustrated in Figure \ref{figure:comparison_fig}. The scattering paths in the analytic framework are vertical, owing to the assumption that encounters are secular (causing no permanent change in the semi-major axis).  The lower panel of Figure \ref{figure:comparison_fig} shows the outcome probability as a function of the projected radius $r_{\bot}$ from the centre of the cluster (see Section \ref{sec:project_avg} for details). The HJ formation probability falls off steeply beyond a few parsecs from the centre, where the stellar density becomes too low and encounters become too infrequent to effectively pump up the planetary eccentricity. The same is true for all of the other outcomes that require dynamical processing (ionisation, tidal disruption and WJ formation).

\begin{figure} 
\includegraphics[width=\linewidth]{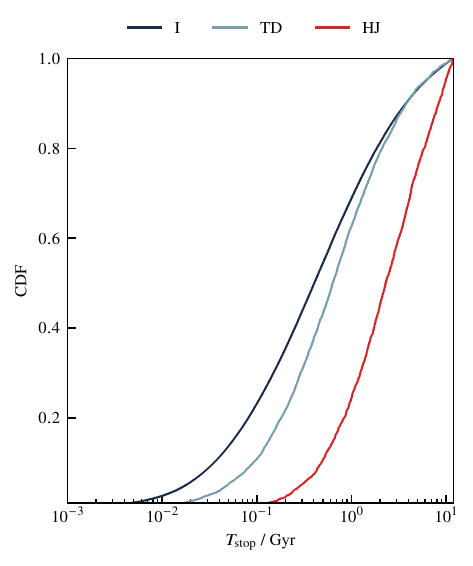}
\caption{The distribution of stopping times for hybrid-model outcomes. HJs only form in significant quantities after a few hundred $\mathrm{Myr}$, when the excited planetary orbits have had sufficiently long to circularise.}
\label{figure:stopping_time_fig}
\end{figure}

\begin{figure} 
\includegraphics[width=\linewidth]{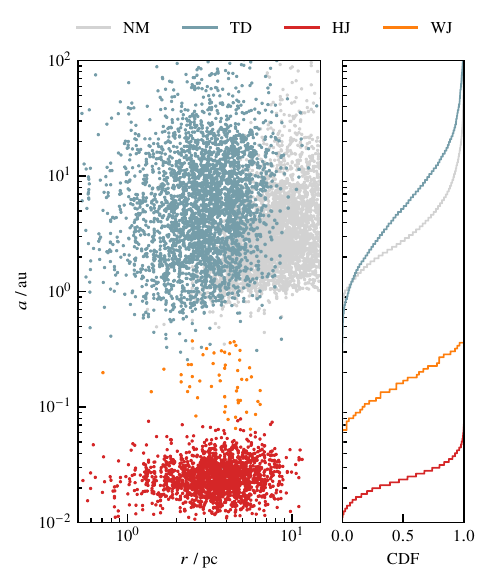}
\caption{\textbf{Left}: The present-day dependence of the final semi-major axis on radial coordinate, $r$, in the cluster, for the hybrid MC model. \textbf{Right}: The cumulative distribution of final semi-major axes.}
\label{figure:semi_major_axis_fig}
\end{figure}

Figure \ref{figure:stopping_time_fig} shows the cumulative density distribution of the stopping time $T_{\mathrm{stop}}$ (see Section \ref{section:stopping_conditions}) for ionisation, tidal disruption and HJ formation outcomes under the hybrid MC model.  Ionisation events occur right from the start of the simulation, as it takes just one strong encounter to unbind a planet from its host star. Tidal disruption events are first observed after just tens of Myr of dynamical evolution, whereas HJ formation events are first observed after hundreds of Myr. HJ formation requires scattering into a critical eccentricity range $e \in [e_{\mathrm{min}}, e_{\mathrm{td}}]$ followed by tidal circularisation. The relatively long timescale of tidal circularisation causes a delay before HJ formation is first observed. In Figure \ref{figure:semi_major_axis_fig} we show the present-day distribution of semi-major axes, $a$, and the radial coordinates of the planetary systems measured from the centre of the cluster, $r$, for the hybrid MC model. The distribution exhibits clear phase space clustering, with HJs at small separations ($0.01 \lesssim a / \mathrm{au} \lesssim 0.1$), WJs at intermediate separations ($0.1 \lesssim a/\mathrm{au} \lesssim 1 \ \mathrm{au}$) and both tidal disruptions and non-migrators beyond $1 \ \mathrm{au}$. Note that there is little difference in these populations as a function of radius within $r \sim 8 \ \mathrm{pc}$, whereas at larger radii the planet population is dominated by non-migrators.

\subsubsection{Projection averaging}
\label{sec:project_avg}
The projected radius $r_{\perp}$ from the centre of the cluster is an observable quantity on the sky-plane. We proceed stochastically by assigning each planetary system an azimuthal angle $\vartheta_i$ relative to the line of sight, sampled uniformly in $\sin{\vartheta_i}$ in the interval $[0, \pi]$. The projected radius is $(r_{\perp})_i = r_i \sin{\vartheta_i}$. Figure \ref{figure:comparison_fig} shows the outcome probability as a function of the projected radius, which has been obtained by normalising the value counts for each outcome in every $r_{\perp}$-bin by the total value count in that bin. We show only the bins containing at least $1,000$ experiments, thereby ensuring that the Poisson error is small ($\sim 0.03$). The probabilities of HJ formation, tidal disruption and ionisation events remain largely constant with projected radius within $r_{\perp}/\mathrm{pc} \sim$ a few parsecs. The sharp drop-off at larger radii is a direct consequence of the steepness of the Plummer density profile beyond the half-mass radius $r_{\mathrm{h}}$ (see Equation \ref{equation:rho}).

\subsubsection{Comparison with observational constraints}

In order to calculate the cluster-wide formation rate of HJs in 47 Tuc, we must make an assumption about the occurrence rate of progenitor cold Jupiter planets. Recently, \citet{FulRos21} applied a hierarchical Bayesian model to RV data from the California Legacy Survey (CLS) to estimate the occurrence rate of giant planets as a function of orbital separation. The per-star occurrence rate for cold Jupiters with semi-major axes between 1-30 au and masses between 30-6000 $M_{\oplus}$ is of the order $N_{\mathrm{mp}} \approx 0.1$ (see Figure 2 of \citet{FulRos21}). The occurrence rates per \textit{cluster star} can therefore be approximated\footnote{The occurrence rate $N_{\mathrm{mp}}$ drops off slightly beyond $a_0 \sim 10 \ \mathrm{au}$, however the fit at large orbital separation is highly uncertain. In any case, the majority of our samples have small orbital separations, so this drop-off does not affect our results.} by 
\begin{equation}
\mathcal{F} \approx 0.1 P. 
\end{equation}
The expected number of HJ detections in a survey of $N$ stars is $\lambda = \mathcal{F}_{\mathrm{HJ}} N \epsilon$, where $\epsilon$ is the average detection efficiency of the survey. Assuming that the number of detected HJs is Poisson distributed, the probability of zero HJ detections is $P(0|\lambda) = e^{-\lambda}$. At 95\% confidence, the upper bound on the HJ occurrence probability implied by a null survey result is therefore $P_{\mathrm{HJ}} \lesssim 3/(N\epsilon)$.

\citet{Gil00}, hereafter G00, performed a transit survey of 34,091 main-sequence stars near the core of 47 Tuc, targeting stars with visual magnitudes $17.1 < V < 21.1$, radii $0.8 < R_{\mathrm{p}}/R_{\mathrm{J}} < 2$ and periods $0.5 < T/\mathrm{days} < 8.3$. The authors expected to find 17 HJs (assuming an occurrence rate of $P_{\mathrm{HJ}} \sim 1\%$ in the field), thereby implying a survey sensitivity of $N\epsilon \sim 1900$. The upper bound on the occurrence rate in the core of 47 Tuc so-obtained is $\mathcal{F}_{\mathrm{HJ}} < 1.6 \times 10^{-3}$. The results of our simulations within $r < 0.5 \ \mathrm{pc}$ (in the inner core) are consistent with this constraint; the hybrid MC model yields $\mathcal{F}_{\mathrm{HJ}}^{<0.5} = 1.2\times 10^{-3}$. This would suggest that the suppression of high-$e$ migration at high densities can indeed explain the lack of HJ detections in the core of 47 Tuc.

In the outskirts of 47 Tuc, the recent MISHAPS survey \citet{cri24} yields an upper bound of $\mathcal{F}_{\mathrm{HJ}} < 3.6 \times 10^{-3}$, when applying the same criteria for planetary-radii and periods as G00. The results of our simulations outside of $r > 8 \ \mathrm{pc}$ are well within this constraint, with the analytic and hybrid MC models both yielding $\mathcal{F}_{\mathrm{HJ}}^{>8} = 4 \times 10^{-5}$. The lower survey sensitivity ($N\epsilon = 830$) results in a weaker observational upper bound in the outskirts when compared to the G00 bound in the core, even though our simulations suggest that we would expect to find almost no HJs in the outskirts.

\subsubsection{Comparison with \citet{Win22}}

The overall HJ formation rate in 47 Tuc implied by the analytic and hybrid MC models are $\mathcal{F}_{\mathrm{HJ}} \approx 3.9 \times 10^{-4}$ and $\mathcal{F}_{\mathrm{HJ}} \approx 5.9 \times 10^{-4}$ respectively. Both of these values are smaller than the analytic estimate $\mathcal{F}_{\mathrm{HJ}} \approx 2.2 \times 10^{-3}$ obtained by \citet{Win22} by an order unity factor. The fact that our analytic estimate does not exactly coincide with the aforementioned value is in part due to our simplified `toy model' of 47 Tuc (see Section \ref{section:toy_model}). The work described in \citet{Win22} employs a full density-evolution model. Since the results of our toy-model experiments (see Table \ref{table:47_tuc_results}) demonstrate that the improved `hybrid' treatment of dynamical encounters leads to an enhancement in $\mathcal{F}_{\mathrm{HJ}}$ compared to analytic predictions, we might hypothesize that running the hybrid MC model on an identical cluster model to \citet{Win22} would yield a \textit{larger} than $\mathcal{F}_{\mathrm{HJ}} \approx 2.2 \times 10^{-3}$. As shown in Figure \ref{figure:test_model_param_plot}, however, the analytic $P_{\mathrm{HJ}}$ can both under-estimate or over-estimate the actual $P_{\mathrm{HJ}}$ depending on the specific values of $n$, $\sigma$ or $a_0$, for example. The overall occurrence rates are quite sensitive to the cluster model, and it is not obvious in the general case whether the relative enhancement in $\mathcal{F}_{\mathrm{HJ}}$ will be comparable to that obtained in this paper.

It is also interesting to note that host stars of planets that are ionised, or at large radii, will be preferentially ejected from the cluster. From the results of Figure \ref{figure:comparison_fig}, we expect that this effect would push our results in the direction of higher $\mathcal{F}_{\mathrm{HJ}}$, towards that of \citet{Win22}.

\subsubsection{Comparison with FFP occurrence rates}
In the hypothetical limit that all field stars originate in clusters (and that their ionized planets end up in the field), the ionisation rate implied by the hybrid  MC model yield $\mathcal{F}_{\mathrm{FFP}} \approx N_{\mathrm{mp}} P_{\mathrm{ion}} \approx 0.062$ free-floating Jupiter-size planets per cluster star. This is an order of magnitude lower than the OGLE-IV $95\%$ upper-bound of $\mathcal{F}_{\mathrm{FFP}}(M \approx M_{\mathrm{J}}) < 0.25$ \citep{Mroz17}. On the other hand, a re-examination of MOA-II data by \citet{Sumi23} suggests a power-law mass function 
\begin{equation}
\frac{dN}{d\log{M}} = 2.18 \left( \frac{M}{8M_{\oplus}} \right)^{-0.96}
\end{equation}
to the FFP population, which at $M=M_{\mathrm{J}}$ evaluates to about $0.064$ FFP per star per dex. Our model could in principle generate the MOA-II results if the field consists of a high fraction of cluster stars along with their ionized planets.

\subsection{Limitations of the model} 

\subsubsection{Post-MS stellar expansion and engulfment}

We have treated the host stars in this work as time-invariant point masses with no stellar evolution. This is a fair assumption for our work, since the majority of constraints, i.e. \citet{Gil00}, are for main sequence stars. In samples containing post-MS stars, a portion of the stellar population will exhaust core hydrogen and evolve onto the red giant branch (RGB) during the $\sim 10 \ \mathrm{Gyr}$ timescale of our simulations. Hydrodynamic simulations by \citet{Lau25} show that once a Jupiter-mass planet is engulfed by a solar-mass star that has evolved to a radius of $R_{\star} = 4 \ R_{\odot}$ ($0.019 \ \mathrm{au}$), ram-pressure drag drives orbital inspiral and total disruption within $t\lesssim 80 \ \mathrm{hr}$. Any HJ whose instantaneous pericentre satisfies $a(1-e) < R_{\star}(t)$ will be removed on a timescale that is effectively instantaneous compared to the timescale of our simulations. The HJ occurrence rate obtained in this work may therefore be regarded as an upper bound on the observable occurrence rate for an arbitrary sample.

\subsubsection{Primordial and dynamical binaries}

Dense clusters contain both primordial and dynamically-assembled stellar binaries.  A recent VLT/MUSE survey of 21\,699 stars in the core of 47~Tuc determines a present–day spectroscopic binary fraction of $(2.4\pm1.0)\%$ \citep{MulHor25}. Even if the initial binary fraction of $0.8 \ M_{\odot}$ stars were an order of magnitude higher than this, as was suggested for field stars by \citet{Moe2017}, stellar perturbation in globular cluster environments would rapidly disrupt soft pairs and harden the survivors \citep{Heggie1975}. Binary hardening preferentially ejects the lowest-mass component, so one wouldn't expect many giant planets to survive the dynamical evolution of a stellar binary. We defer a detailed treatment to a future study.

\subsubsection{Planetary companions}

We have considered only single-planet systems, yet multi-planet architectures are common among gas-giant hosts. \citet{Ben24} showed that the presence of a $1 \ \mathrm{M}_{\mathrm{J}}$ planet at $a = 30 \ \mathrm{au}$ roughly triples the HJ formation probability $P_{\mathrm{HJ}}$ in low-density environments by enabling secular angular-momentum-deficit (AMD) transfer from outer to inner planets following stellar perturbations. In denser regions, however, the encounter rate exceeds the secular precession frequency of wide companions, and hence direct stellar kicks dominate the eccentricity evolution.

\section{Conclusions} \label{section:conclusions}

In this work, we have introduced an efficient algorithm for computing the long-term eccentricity evolution of planetary systems in stellar clusters. The effect of dynamical encounters outside of the tidal and slow regime has been accurately modelled by direct N-body integration. Our results indicate that the probability of HJ formation peaks at stellar densities of order $n \sim 10^4 \ \mathrm{pc}^{-3}$, in agreement with \citet{HamTre17}, where the rate of encounters is sufficiently high to drive efficient eccentricity diffusion but not high enough to disrupt the majority of circularising planets. HJ formation was shown to peak at low semi-major axes ($a_0 \sim 1 \ \mathrm{au}$) and low velocity dispersions.  

We applied our model to planetary systems dispersed within a simple, time-dependent Plummer profile for 47 Tuc. The HJ formation rate per cluster star in our hybrid MC model ($\mathcal{F}_{\mathrm{HJ}} \approx 5.9 \times 10^{-4}$) was enhanced by $51$ per cent compared to the analytic MC model ($\mathcal{F}_{\mathrm{HJ}} \approx 3.9 \times 10^{-4}$). As shown in Section \ref{section:eccentricity-diffusion}, direct N-body integration of encounters outside of the tidal and slow regime leads to an increased rate of eccentricity diffusion. We demonstrated in Figure \ref{figure:test_model_param_plot} that the hybrid-model HJ formation rates for fixed environmental parameters $(n, \sigma)$ exceeded the analytic prediction at intermediate densities $n / \mathrm{pc}^{-3} \in [10^3, 10^5]$, which are typical of the core of 47 Tuc.

We expect that this pipeline will be useful for interpreting results from future observational surveys, such as the Roman Galactic Bulge Time Domain Survey \citep{Gru23}, should it target globular clusters (and particularly those with high metallicities). The detection or non-detection of HJs in future surveys of 47 Tuc will provide rich insight into planet formation pathways. In particular, obtaining a robust model for HJ formation would allow us to invert the present day statistics of HJs in globular clusters and constrain the birth population of Jovian-size gas giants at wider radii (5-10 au). 

\section*{Acknowledgements}

This work made use of the \texttt{REBOUND} N-body library \citep{ReiLiu12}. AJW has received funding from the European Union’s Horizon 2020 research and innovation programme under the Marie Skłodowska-Curie grant agreement No 101104656.

%%%%%%%%%%%%%%%%%%%%%%%%%%%%%%%%%%%%%%%%%%%%%%%%%%
\section*{Data Availability}

The code used for this work is available at \url{https://github.com/James-Wirth/HotJupiter}, and the simulation configuration files are provided within the repository to enable reproducibility. Data generated during this study are available upon reasonable request from the corresponding author.



\bibliographystyle{mnras}
\bibliography{example} % if your bibtex file is called example.bib

%\begin{thebibliography}{99}
%\end{thebibliography}

%%%%%%%%%%%%%%%%%%%%%%%%%%%%%%%%%%%%%%%%%%%%%%%%%%

%%%%%%%%%%%%%%%%% APPENDICES %%%%%%%%%%%%%%%%%%%%%

\appendix

\section{Analytic approximation}
\label{section:analytic_approximation}

The analytic approximation for the leading order eccentricity excitation due to a perturbing stellar encounter is
\begin{align}  \label{equation:eccentricity_excitation}
\epsilon = \alpha y \left(\frac{a}{R_{\mathrm{peri}}}\right)^{3/2}\{ \Theta_1 \chi + [\Theta_2 + \Theta_3]\psi \}
\end{align}
The auxiliary functions $\alpha$, $y$, $\chi$, $\psi$ and angular terms $\Theta_i$ are
\[
\begin{array}{rl}
\alpha & = -\frac{15}{4}(1+e_{\mathrm{pert}})^{-3/2} \\
y & = e \sqrt{1-e^2} 
      \frac{m_{\mathrm{pert}}}{\sqrt{m_{\star}(1+q) m_{\mathrm{tot}}}} \\
\chi & = \cos^{-1}(-1/e_{\mathrm{pert}}) + \sqrt{e_{\mathrm{pert}}^2-1} \\
\psi & = \frac{1}{3} (e_{\mathrm{pert}}^2-1)^{3/2} e_{\mathrm{pert}}^{-2} \\[1em]
\Theta_1 & = \sin^2{i} \sin{2\Omega} \\
\Theta_2 & = (1+\cos^2{i})\cos{2\omega} \sin{2\Omega} \\
\Theta_3 & = 2\cos{i} \sin{2\omega} \cos{2\Omega}
\end{array}
\]
The starting point in the derivation of this result is the disturbing function $\varphi$ given in Equation \ref{equation:multipole}. To quadrupole order, the disturbing force $\mathbf{F} = \nabla_{\mathbf{R}} \varphi$ is
\begin{equation}
\mathbf{F} = \frac{Gm_{\mathrm{pert}}r}{R^3}(3\nu \hat{\mathbf{R}} - \hat{\mathbf{r}})
\end{equation}
where $\nu := \mathbf{r}\cdot\mathbf{R}/(rR)$. This expression holds to good approximation in the tidal regime (Equation \ref{equation:tidal-condition}). \citet{Heggie1996} consider the evolution of the eccentricity vector, $\mathbf{e}$, of the planetary orbit, defined by 
\begin{equation} \label{equation:ecc_vector}
Gm_{\star}(1+q)\mathbf{e} := \dot{\mathbf{r}} \times (\mathbf{r} \times \dot{\mathbf{r}}) - \hat{\mathbf{r}}
\end{equation}
By inserting the equation of motion $\ddot{\mathbf{r}} = -Gm_{\star}(1+q)\mathbf{r}/r^3 + \mathbf{F}$ into the time derivative of equation \ref{equation:ecc_vector}, we obtain
\begin{equation}
Gm_{\star}(1+q)\dot{\mathbf{e}} = 2(\mathbf{F}\cdot \dot{\mathbf{r}})\mathbf{r} - (\mathbf{r} \cdot \dot{\mathbf{r}})\mathbf{F} - (\mathbf{F}\cdot{\mathbf{r}})\dot{\mathbf{r}}
\end{equation}
The key insight involves averaging this equation over the planetary orbit, assuming that the period is short compared to the timescale of the perturbing flyby. In the slow regime (Equation \ref{equation:slow-condition}), the vector $\mathbf{R}$ can be treated as a constant during the time-averaging. Equation \ref{equation:eccentricity_excitation} is obtained after a series of tedious, but straightforward, integration steps.

\begin{figure} 
\includegraphics[width=\linewidth]{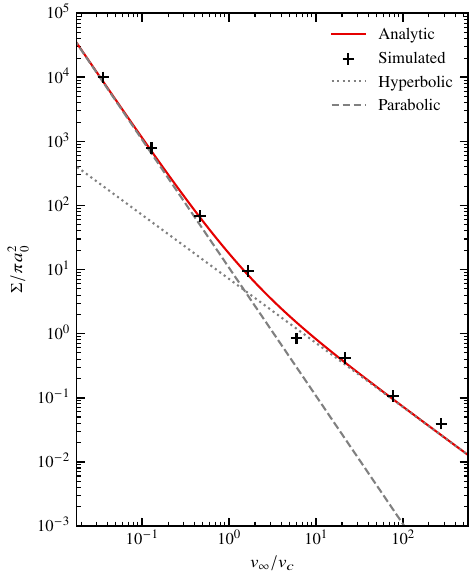}
\caption{The perturbation cross-section for encounters yielding $|\epsilon| > \epsilon^{*} = 0.05$ is plotted as a function of $v_{\infty}/v_{\mathrm{c}}$. For weak encounters ($\epsilon^{*} \rightarrow 0$), the analytic prediction is in good agreement with the simulated results in both the parabolic and hyperbolic regimes. We have fixed $\Omega = i = \omega = 1 \ \mathrm{rad}$, $e_0 = 0.9$, $m_{\star} = m_{\mathrm{pert}} = M_{\odot}$ and $m_{\mathrm{p}} = 10^{-9} \ M_{\odot} \approx 0$.} 
\label{figure:cross_section}
\end{figure}

We follow \citet{Fre04} in defining the perturbation cross-section $\Sigma(v_{\infty};\epsilon^{*})$ in terms of the critical impact parameter $b^{*}$ such that $|\epsilon| > \epsilon^{*}$ for all $b < b^{*}$. For the simulated results, $\epsilon := \langle \epsilon \rangle_{\boldsymbol{\phi}}$ is taken to be the phase-averaged eccentricity excitation.
\begin{equation}
\Sigma(v_{\infty};\epsilon^{*}) = \frac{1}{4\pi^2} \oint \mathrm{d}\Omega \oint \mathrm{d}\omega \oint \mathrm{d}i \ \frac{1}{2} \sin{i} \int_0^{b^{*}} \mathrm{d}b \ 2 \pi b
\end{equation}
In Figure \ref{figure:cross_section} the perturbation cross-section obtained via direct N-body integration is compared to the analytic cross-section \citep{Heggie1996} for a small threshold excitation $\epsilon^{*} = 0.05$. We observed that the analytic approximation converges to the simulated cross-section as $\epsilon^{*} \rightarrow 0$, corresponding to the limit of weak encounters. The asymptotic relative speed is expressed as a fraction of the characteristic binary orbital speed $v_{\mathrm{c}} := (Gm_{\star}(1+q)/a)^{1/2}$. The analytic cross-section has been expressed as the sum of two independent contributions: a hyperbolic component $\Sigma^{(\mathrm{hyp})} \sim v_{\infty}^{-1}$ and a parabolic component $\Sigma^{(\mathrm{para})} \sim v_{\infty}^{-2}$ \citep{Win22}.

%%%%%%%%%%%%%%%%%%%%%%%%%%%%%%%%%%%%%%%%%%%%%%%%%%

% Don't change these lines
\bsp	% typesetting comment
\label{lastpage}
\end{document}